%% file: main_arxiv.tex
\documentclass[
reprint,
superscriptaddress,
nofootinbib,
aps,
prl
]{revtex4-2}

\usepackage[version=3]{mhchem}
\usepackage{todonotes}
\usepackage{glossaries}
\usepackage{chemformula}
\usepackage{booktabs}
\usepackage{multirow}
\usepackage{bbm}
\usepackage{amsmath}
\usepackage{amssymb}
\usepackage{enumitem}

\newcommand{\algo}[0]{\textit{AdsorbML}}

\newcommand{\modelsurl}[0]{\url{https://github.com/Open-Catalyst-Project/ocp/blob/main/MODELS.md}}
\newcommand{\code}[0]{\url{https://github.com/Open-Catalyst-Project/AdsorbML}}
\newcommand{\configs}[0]{\url{https://github.com/Open-Catalyst-Project/AdsorbML/tree/main/configs}}

\newacronym{DFT}{DFT}{Density Functional Theory}
\newacronym{DFTHR}{DFT-Heur+Rand}{DFT-Heuristic+Random}
\newacronym{DFTH}{DFT-Heur}{DFT-Heuristic-Only}
\newacronym{SI}{SI}{Supporting Information}
\newacronym{PES}{PES}{Potential Energy Surface}
\newacronym{VASP}{VASP}{\textit{Vienna Ab initio Simulation Package}}
\newacronym{RPBE}{RPBE}{revised Perdew-Burke-Ernzerhof}
\newacronym{PAW}{PAW}{projector augmented wave}
\newacronym{S2EF}{\textit{S2EF}}{Structure to Energy and Forces}
\newacronym{IS2RE}{\textit{IS2RE}}{Initial Structure to Relaxed Energy}
\newacronym{ASE}{ASE}{Atomic Simulation Environment}
\newacronym{MAE}{MAE}{Mean Absolute Error}
\newacronym{CO2RR}{\ch{CO2}RR}{\ch{CO2} Reduction Reaction}
\newacronym{ID}{ID}{In-Domain}
\newacronym{OOD}{OOD}{Out-of-Domain}
\newacronym{OC20}{OC20}{Open Catalyst 2020 Dataset}
\newacronym{OC22}{OC22}{Open Catalyst 2022 Dataset}
\newacronym{ML}{ML}{machine learning}
\newacronym{GNNs}{GNNs}{Graph Neural Networks}
\newacronym{GNN}{GNN}{Graph Neural Network}
\newacronym{MD}{MD}{\textit{ab initio} Molecular Dynamics}
\newacronym{OC20D}{OC20-Dense}{Open Catalyst 2020 - Dense Dataset}
\newacronym{DFTB}{DFTB}{Density Functional based Tight Binding}
\newacronym{OCP}{OCP}{Open Catalyst Project}
\newacronym{SC}{SC}{self-consistency}
\newacronym{DBSCAN}{DBSCAN}{density-based spatial clustering of applications with noise}

\newcommand{\fair}{Fundamental AI Research (FAIR), Meta AI, Meta}
\newcommand{\coa}{$^{*}$}
\newcommand{\cmu}{Department of Chemical Engineering, Carnegie Mellon University}
\newcommand{\scott}{Scott Institute for Energy Innovation, Carnegie Mellon University}

\begin{document}

\author{Janice Lan\coa}
\affiliation{\fair}
\author{Aini Palizhati\coa}
\affiliation{\cmu}
\author{Muhammed Shuaibi\coa}
\affiliation{\fair}
\author{Brandon M. Wood\coa}
\affiliation{\fair}
\author{Brook Wander}
\affiliation{\cmu}
\author{Abhishek Das}
\affiliation{\fair}
\author{Matt Uyttendaele}
\affiliation{\fair}
\author{C. Lawrence Zitnick$^{\dagger}$}
\affiliation{\fair}
\author{Zachary W. Ulissi$^{\dagger}$}
\affiliation{\cmu}
\affiliation{\scott}

\def\thefootnote{$*$}\footnotetext{\textbf{Equal Contribution}}\def\thefootnote{\arabic{footnote}}
\def\thefootnote{$\dagger$}\footnotetext{\textbf{Corresponding authors\\C.L.Z., email: \url{zitnick@meta.com}\\Z.W.U., email: \url{zulissi@andrew.cmu.edu}}\\}\def\thefootnote{\arabic{footnote}}

\title[]{AdsorbML: A Leap in Efficiency for Adsorption Energy Calculations using Generalizable Machine Learning Potentials}

\begin{abstract}\input{sections/abstract.tex}
\end{abstract}
\maketitle
\clearpage

\section{Introduction}
\input{sections/intro.tex}
\subsection{Related Work}
\input{sections/related.tex}
\input{sections/paper_contributions.tex}
\input{sections/results.tex} 
\input{sections/discussion.tex}
\input{sections/methods.tex}

\section{Data Availability}
The full open dataset is provided at \code{}.

\section{Code Availability}
All accompanying code is provided at \code{}.

\section{Author Contributions}
\input{sections/author_contributions.tex}

\section{Competing Interests}
The authors declare no competing interests.

\renewcommand{\figurename}{Supplementary Figure}
\renewcommand{\tablename}{Supplementary Table}
\setcounter{table}{0}
\setcounter{figure}{0}
\input{sections/si_description}

\clearpage
\bibliography{main.bib}
\clearpage

\onecolumngrid
\input{sections/supplementary.tex}
\clearpage
\input{sections/changelog.tex}

\end{document}

%% file: sections/abstract.tex
Computational catalysis is playing an increasingly significant role in the design of catalysts across a wide range of applications. A common task for many computational methods is the need to accurately compute the adsorption energy for an adsorbate and a catalyst surface of interest. Traditionally, the identification of low energy adsorbate-surface configurations relies on heuristic methods and researcher intuition. As the desire to perform high-throughput screening increases, it becomes challenging to use heuristics and intuition alone. In this paper, we demonstrate machine learning potentials can be leveraged to identify low energy adsorbate-surface configurations more accurately and efficiently. Our algorithm provides a spectrum of trade-offs between accuracy and efficiency, with one balanced option finding the lowest energy configuration 87.36\% of the time, while achieving a $\sim$2000x speedup in computation. To standardize benchmarking, we introduce the Open Catalyst Dense dataset containing nearly 1,000 diverse surfaces and $\sim$100,000 unique configurations.

%% file: sections/intro.tex
The design of novel heterogeneous catalysts plays an essential role in the synthesis of everyday fuels and chemicals. To accommodate the growing demand for energy while combating climate change, efficient, low-cost catalysts are critical to the utilization of renewable energy~\cite{norskov_book, Chanussot2021, Dumesic, zitnick2020introduction}.  Given the enormity of the material design space, efficient screening methods are highly sought after \cite{zitnick2020introduction, choudhary2022recent, wen2022deep,wei2019machine}. Computational catalysis offers the potential to screen vast numbers of materials to complement more time- and cost- intensive experimental studies.

A critical task for many first-principles approaches to heterogeneous catalyst discovery is the calculation of adsorption energies. The adsorption energy is the energy associated with a molecule, or adsorbate, interacting with a catalyst surface. Adsorbates are often selected to capture the various steps, or intermediates, in a reaction pathway (e.g. *\ch{CHO} in \ch{CO2} reduction). Adsorption energy is calculated by finding the adsorbate-surface configuration that minimizes the structure's overall energy. Thus, the adsorption energy is the global minimum energy across all potential adsorbate placements and configurations. These adsorption energies are the starting point for the calculation of the free energy diagrams to determine the most favorable reaction pathways on a catalyst surface ~\cite{ulissi2017address}. It has been demonstrated that adsorption energies of reaction intermediates can be powerful descriptors that correlate with experimental outcomes such as activity or selectivity \cite{tran2018active, zhong2020accelerated, liu2017understanding, norskov2005trends, wan2022machine}. This ability to predict trends in catalytic properties from first-principles is the basis for efficient catalyst screening approaches \cite{She2017, norskov_book}. 

\begin{figure*}[t]
    \centering
    \includegraphics[width=\textwidth]{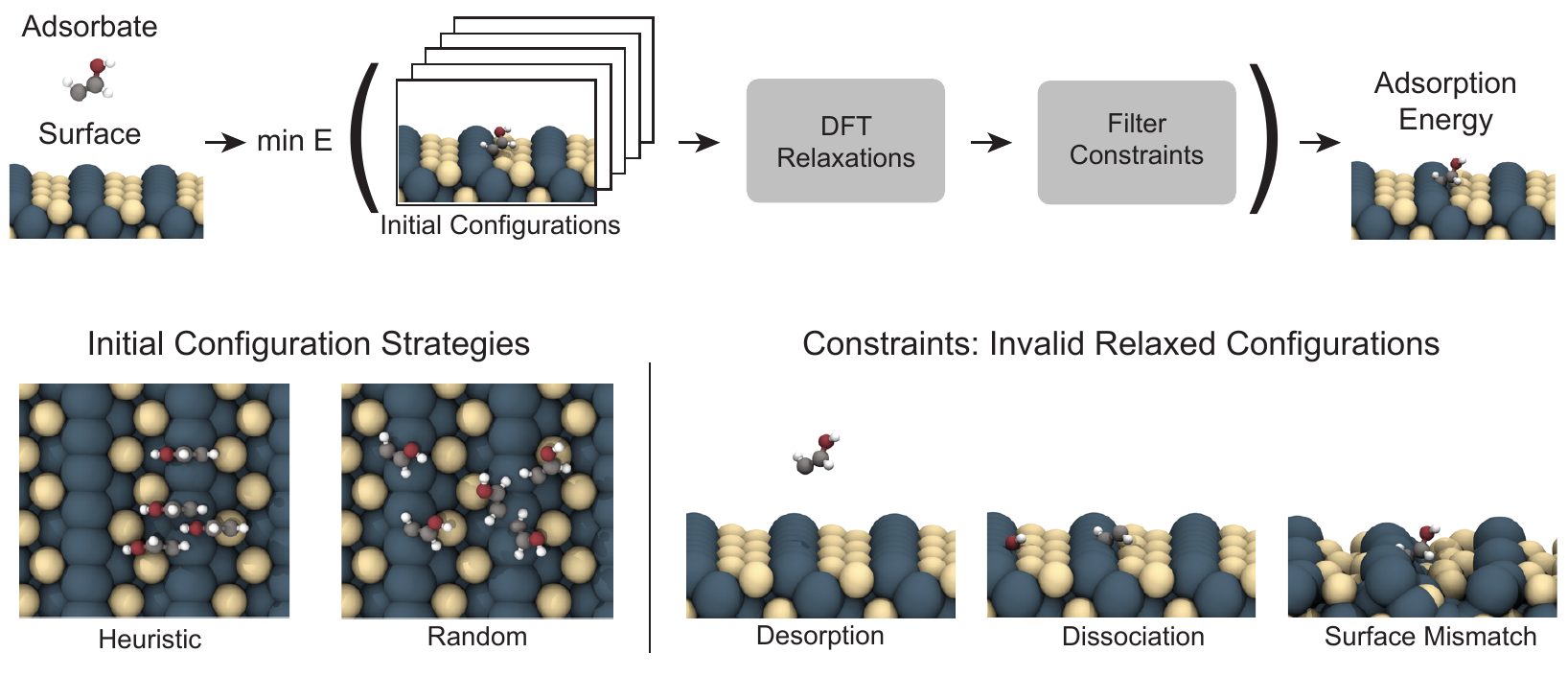}
    \caption{An overview of the steps involved in identifying the adsorption energy for an adsorbate-surface combination. First, an adsorbate and surface combination are selected, then numerous configurations are enumerated heuristically and/or randomly. For each configuration, \gls{DFT} relaxations are performed and systems are filtered based on physical constraints that ensure valid adsorption energies (i.e. desorption, dissociation, surface mismatch). The minimum energy across all configurations is identified as the adsorption energy.}
    \label{fig:ad_e_overview}
\end{figure*}

Finding the adsorption energy presents a number of complexities. There are numerous potential binding sites for an adsorbate on a surface and for each binding site there are multiple ways to orient the adsorbate (see bottom-left in Figure \ref{fig:ad_e_overview}). When an adsorbate is placed on a catalyst's surface, the adsorbate and surface atoms will interact with each other. To determine the adsorption energy for a specific adsorbate-surface configuration, the atom positions need to be relaxed until a local energy minimum is reached. \gls{DFT}~\cite{hohenberg1964inhomogeneous, kohn1965self, Sholl2009} is the most common approach to performing this adsorbate-surface relaxation. DFT first computes a single-point calculation where the output is the system's energy and the per-atoms forces. A relaxation then performs a local optimization where per-atom forces are iteratively calculated with DFT and used to update atom positions with an optimization algorithm (e.g. conjugate gradient~\cite{cg}) until a local energy minimum is found. To find the global minimum, a strategy for sampling adsorbate-surface configurations and/or a technique such as minima hopping~\cite{Peterson2014, goedecker2004minima} for overcoming energy barriers during optimization is required. 

Adsorption energy ($\Delta E_{\text{ads}}$) is calculated as the energy of the adsorbate-surface ($E_{\text{sys}}$) minus the energy of the clean surface (i.e. slab) ($E_{\text{slab}}$) and the energy of the gas phase adsorbate or reference species ($E_{\text{gas}}$), as defined by Chanussot, et al. and detailed in the \gls{SI}.~\cite{Chanussot2021, zitnick2020introduction}
\begin{equation}
\Delta E_{\text{ads}} = E_{\text{sys}} - E_{\text{slab}} - E_{\text{gas}}
\end{equation}
Relaxed adsorbate-surface structures must respect certain desired properties in order for their adsorption energy to be both accurate and valid. One example of a constraint is the adsorbate should not be desorbed, i.e., float away, from the surface in the final relaxed structure (Figure \ref{fig:ad_e_overview} bottom-right).  Additionally, if the adsorbate has multiple atoms it should not dissociate or break apart into multiple adsorbates because it would no longer be the adsorption energy of the molecule of interest ~\cite{Peterson2014, jung2022machine}. Similarly, if the adsorbate induces significant changes in the surface compared to the clean surface, the $E_{\text{slab}}$ reference would create a surface mismatch. It is important to note that if a relaxed structure breaks one of these constrains it does not necessarily mean the relaxation was inaccurate; these outcomes do arise but they lead to invalid or inaccurate adsorption energies as it has been defined.

Identifying the globally optimal adsorbate-surface configuration has historically relied on expert intuition or more recently heuristic approaches. Intuition and trial and error can be used for one-off systems of interest but it does not scale to large numbers of systems. Commonly used heuristics are often based on surface symmetry~\cite{ong2013python, catkit}. These methods have been used successfully in past descriptor-based studies~\cite{Andersson2006, Bligaard2004, Studt2008, Nilekar2011, zhong2020accelerated, tran2018active}. More recently, a graph-based method has been used to identify unique adsorbate-surface configurations~\cite{deshpande2020graph}. Nevertheless, as the complexity of the surfaces and adsorbates increase, the challenge of finding the lowest energy adsorbate-surface configuration grows substantially. This is especially challenging when the adsorbate is flexible, having multiple configurations of its own, such that there are many effective degrees of freedom in the system. 

While \gls{DFT} offers the ability to accurately estimate atomic forces and energies, it is computationally expensive, scaling $O(N^3)$ with the number of electrons. Evaluating a single adsorbate-surface configuration with a full \gls{DFT} relaxation can take $\sim$24 hours to compute~\cite{Chanussot2021, tran2022open}. Since numerous configurations are typically explored to find the adsorption energy, all the \gls{DFT} calculations involved can take days or even weeks. Hypothetically, if one were to brute force screen 100,000 materials from the Materials Project database~\cite{jain2013commentary} for \gls{CO2RR} using 5 adsorbate descriptors, $\sim$90 surfaces/material, and $\sim$100 sites/surface, one would need $\sim$4.5 billion CPU-days of compute, an intractable problem for even the world's largest supercomputers. To significantly reduce the required computation, a promising approach is to accelerate the search of lowest energy adsorbate-surface configurations with machine learned potentials.

Recently, \gls{ML} potentials for estimating atomic forces and energies have shown significant progress on standard benchmarks while being orders of magnitude faster than DFT ~\cite{schutt2017schnet, klicpera2020directional, klicpera2020fast, gasteiger2022graph, zitnick2022spherical, Chanussot2021, chmiela2017machine}. While \gls{ML} accuracies on the large and diverse \gls{OC20} dataset have improved to 0.3 eV for relaxed energy estimation, an accuracy of 0.1 eV is still desired for accurate screening ~\cite{kolluru2022open}. This raises the question of whether a hybrid approach that uses both \gls{DFT} and \gls{ML} potentials can achieve high accuracy while maintaining efficiency. 

Assessing the performance of new methods for finding low energy adsorbate-surface configurations is challenging without standardized validation data. It is common for new methods to be tested on a relatively small number of systems, which makes generalization difficult to evaluate~\cite{Peterson2014, Chang2021, deshpande2020graph, Chan2019, Fang2021}. While \gls{OC20} contains $O(1M)$ ``adsorption energies", it did not sample multiple configurations per adsorbate-surface combination meaning the one configuration that was relaxed is unlikely to be the global minimum. This makes \gls{OC20} an inappropriate dataset for finding the minimum binding energy~\cite{Chanussot2021}. To address this issue, we introduce the \gls{OC20D}. \gls{OC20D} includes two splits - a validation and test set. The validation set is used for development; and the test set for reporting performance. Each split consists of approximately 1,000 unique adsorbate-surface combinations from the validation and test sets of the \gls{OC20} dataset. No data from \gls{OC20D} is used for training.
To explore the generalizability of our approach, we take $\sim$250 combinations from each of the four \gls{OC20} subsplits - \gls{ID}, \gls{OOD}-Adsorbate, \gls{OOD}-Catalyst, and \gls{OOD}-Both. For each combination, we perform a dense sampling of initial configurations and calculate relaxations using \gls{DFT} to create a strong baseline for evaluating estimated adsorption energies.

We propose a hybrid approach to estimating adsorption energies that takes advantage of the strengths of both \gls{ML} potentials and \gls{DFT}. We sample a large number of potential adsorbate configurations using both heuristic and random strategies and perform relaxations using \gls{ML} potentials. The best-$k$ relaxed energies can then be refined using single-point \gls{DFT} calculations or with full \gls{DFT} relaxations. Using this approach, the appropriate trade-offs may be made between accuracy and efficiency. 

%% file: sections/related.tex
Considerable research effort has been dedicated to determining the lowest energy adsorbate-surface configuration through improvement of initial structure generation and global optimization strategies ~\cite{Peterson2014, deshpande2020graph, Chang2021, Fang2021, Chan2019, jung2022machine, xu2022predicting}. Peterson~\cite{Peterson2014} adopted the minima-hopping method and developed a global optimization approach that preserves adsorbate identity using constrained minima hopping. However, the method relies entirely on \gls{DFT} to perform the search, still making it computationally expensive. More recently, Jung et al.~\cite{jung2022machine} proposed an active learning workflow where a gaussian process is used to run constrained minima hopping simulations. Structures generated by their simulations are verified by \gls{DFT} and iteratively added to the training set until model convergence is achieved. The trained model then runs parallel constrained minima hopping simulations, a subset is refined with \gls{DFT}, and the final adsorption energy identified. We note that prior attempts to use machine learning models to accelerate this process have typically relied on bespoke models for each adsorbate/catalyst combination, which limits broader applicability ~\cite{ulissi2017machine, ghanekar2022adsorbate}. One possibility to greatly expand the versatility of these methods while continuing to reduce the human and computational cost is using generalizable machine learning potentials to accelerate the search for low energy adsorbate-surface configurations.\\

%% file: sections/paper_contributions.tex
The contributions of this work are three-fold:
\begin{itemize}
    \item We propose the \algo{} algorithm to identify the adsorption energy under a spectrum of accuracy-efficiency trade-offs.
    \item We develop the \acrfull{OC20D} to benchmark the task of adsorption energy search.
    \item We benchmark literature \gls{GNN} models on \gls{OC20D} using the proposed \algo{} algorithm; identifying several promising models well-suited for practical screening applications. 
\end{itemize}

%% file: sections/results.tex
\section{Results}
\subsection{OC20-Dense Evaluation}
\input{tables/ml_success_analysis.tex}
To evaluate methods for computing adsorption energies, we present the \acrfull{OC20D} that closely approximates the ground truth adsorption energy by densely exploring numerous configurations for each unique adsorbate-surface system. Each \gls{OC20D} split comprises $\sim$1,000 unique adsorbate-surface combinations spanning 74 adsorbates, 800+ inorganic bulk crystal structures, and a total of $80,000+$ heuristically and randomly generated configurations. A summary of the two splits are provided in Table~\ref{tab:oc20dense-splits}. The dataset required $\sim$4 million CPU-hrs of compute to complete. A more detailed discussion on \gls{OC20D} can be found in the Methods section.

We report results on a wide range of \gls{GNN}s previously benchmarked on \gls{OC20} to evaluate the performance of existing models on \gls{OC20D}. These include SchNet~~\cite{schutt2017schnet}, DimeNet++~~\cite{klicpera2020fast, klicpera2020directional}, PaiNN~~\cite{schutt2021equivariant}, GemNet-OC~\cite{gasteiger2022graph}, GemNet-OC-MD~\cite{gasteiger2022graph}, GemNet-OC-MD-Large~\cite{gasteiger2022graph}, SCN-MD-Large~\cite{zitnick2022spherical}, and eSCN-MD-Large~\cite{passaro2023reducing} where MD corresponds to training on \gls{OC20} and its accompanying \gls{MD} dataset. Models were not trained as part of this work; trained models were taken directly from previously published work and can be found at \modelsurl{}. Of the models, (e)SCN-MD-Large and GemNet-OC-MD-Large are currently the top performers on both \gls{OC20} and \gls{OC22}. Exploring the extent these trends hold for \gls{OC20D} will be important to informing how well progress on \gls{OC20} translates to more important downstream tasks like the one presented here.

\input{tables/splits}

Ideally, the ground truth for \gls{OC20D} would be the minimum relaxed energy over all possible configurations for each adsorbate-surface system. Since the number of possible configurations is combinatorial, the community has developed heuristic approaches to adsorbate placement on a catalyst surface \cite{catkit, ong2013python}. When evaluating only heuristic configurations, we refer to this as \gls{DFTH}. To add to the configuration space, we also uniformly sample sites on the surface at random with the adsorbate placed on each of those sites with a random rotation along the z-axis and slight wobble around the x and y axis. When evaluating against both heuristic and random configurations, we refer to this as \gls{DFTHR}. Although computationally more expensive, this benchmark provides a more thorough search of configurations and a more accurate estimate of the adsorption energies than using only heuristic configurations, a common baseline used by the community. More details on the two benchmarks can be found in the Methods section.

\subsection{\gls{ML} Relaxations}
We explore to what extent \gls{ML} predictions can find the adsorption energy within a threshold of the \gls{DFT} minimum energy, or lower. While a perfect \gls{ML} surrogate to \gls{DFT} will only be able to match \gls{DFT}, small errors in the forces and optimizer differences have the potential to add noise to relaxations and result in configurations previously unexplored~\cite{schaarschmidt2022learned}. For each model, relaxations are performed on an identical set of adsorbate configurations. Initial configurations are created based off heuristic strategies commonly used in the literature~\cite{catkit, ong2013python} and randomly generated configurations on the surface. \gls{ML}-driven relaxations are run on all initial configurations; systems not suitable for adsorption energy calculations due to physical constraints are removed, including dissociation, desorption, and surface mismatch. An in-depth discussion on relaxation constraints can be found in the Methods section.

When evaluating performance, we define success as finding an adsorption energy within an acceptable tolerance (0.1 eV in this work~\cite{Chanussot2021, kolluru2022open, schaarschmidt2022learned}) or lower of the \gls{DFT} adsorption energy in \gls{OC20D}. Note that the ground truth adsorption energies in \gls{OC20D} are an upper bound, since it is possible that a lower adsorption energy may exist. When evaluating \gls{ML} predicted adsorption energies, the results must be verified using a single-point DFT calculation, since an evaluation metric without a lower bound could be easily gamed by predicting low energies (see \gls{SI}). To reliably evaluate \gls{ML} we consider an \gls{ML} adsorption energy successful if its within 0.1 eV of the \gls{DFT} adsorption energy or lower, \textbf{and} a corresponding \gls{DFT} single-point evaluation of the predicted \gls{ML} structure is within 0.1 eV of the predicted \gls{ML} energy. This ensures that a \gls{ML} prediction not only found a low adsorption energy but is accurate and not artificially inflated. Results are reported in Table \ref{tab:ml-success}, where top \gls{OC20} models including eSCN-MD-Large and GemNet-OC-MD-Large achieve success rates of 56.52\% and 48.03\%, respectively. Energy \gls{MAE} between \gls{ML} and \gls{DFT} adsorption energies are also reported in Table \ref{tab:ml-success}, correlating well with success rates and \gls{OC20} \gls{S2EF} metrics. 

\begin{figure*}[ht]
    \centering
    \includegraphics[width=\textwidth]{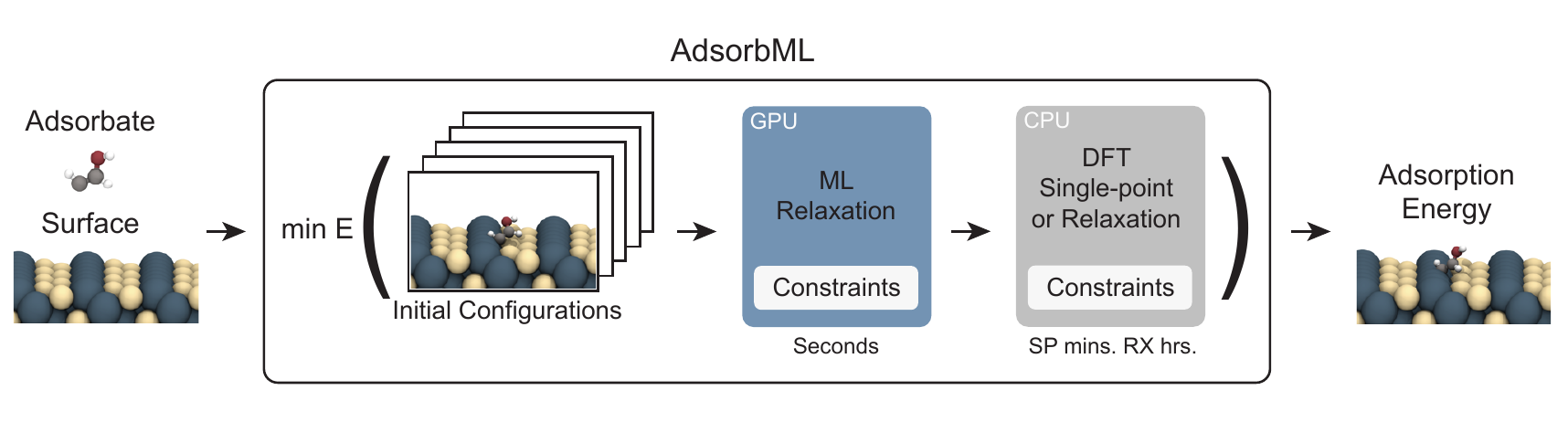}
    \caption{The \algo{} algorithm. Initial configurations are generated via heuristic and random strategies. \gls{ML} relaxations are performed on GPUs and ranked in order of lowest to highest energy. The best $k$ systems are passed on to \gls{DFT} for either a single-point (SP) evaluation or a full relaxation (RX) from the \gls{ML} relaxed structure. Systems not satisfying constraints are filtered at each stage a relaxation is performed. The minimum is taken across all \gls{DFT} outputs for the final adsorption energy.}
    \label{fig:adsorbml_overview}
\end{figure*}
 
While the current state of models have made incredible progress~\cite{kolluru2022open}, higher success rates are needed for everyday practitioners. In a high-throughput setting where successful candidates go on to more expensive analyses or even experimental synthesis, a success rate of $\sim$50\% could result in a substantial waste of time and resources studying false positives. As model development will continue to help improve metrics, this work explores hybrid \gls{ML}+\gls{DFT} strategies to improve success rates at the cost of additional compute.

\subsection{\algo{} Algorithm}
We introduce the \algo{} algorithm to use \gls{ML} to accelerate the adsorbate placement process (Figure \ref{fig:adsorbml_overview}). For each model, we explore two strategies that incorporate \gls{ML} followed by \gls{DFT} calculations to determine the adsorption energy. We note that this strategy is general and can be used with any initial configuration algorithm.

In both approaches the first step is to generate \gls{ML} relaxations. However, rather than taking the minimum across \gls{ML} relaxed energies, we rank the systems in order of lowest to highest energy. The best $k$ systems with lowest energies are selected and (1) \gls{DFT} single-point calculations are done on the corresponding structures (ML+SP) or (2) \gls{DFT} relaxations are performed from \gls{ML} relaxed structures (ML+RX). The first strategy aims to get a more reliable energy measurement of the \gls{ML} predicted relaxed structure, while the second treats \gls{ML} as a pre-optimizer with \gls{DFT} completing the relaxation. By taking the $k$ lowest energy systems, we provide the model with $k$ opportunities to arrive at an acceptably accurate adsorption energy. As we increase $k$, more \gls{DFT} compute is involved, but compared to a full \gls{DFT} approach, we still anticipate significant savings. The adsorption energy for a particular system is obtained by taking the minimum of the best $k$ \gls{DFT} follow-up calculations. 

In both strategies, \gls{ML} energies are used solely to rank configurations, with the final energy prediction coming from a \gls{DFT} calculation. While computationally it would be ideal to fully rely on \gls{ML}, the use of \gls{DFT} both improves accuracy and provides a verification step to bring us more confidence in our adsorption energy predictions. 

\begin{figure*}[t]
    \centering
    \includegraphics[width=0.9\textwidth]{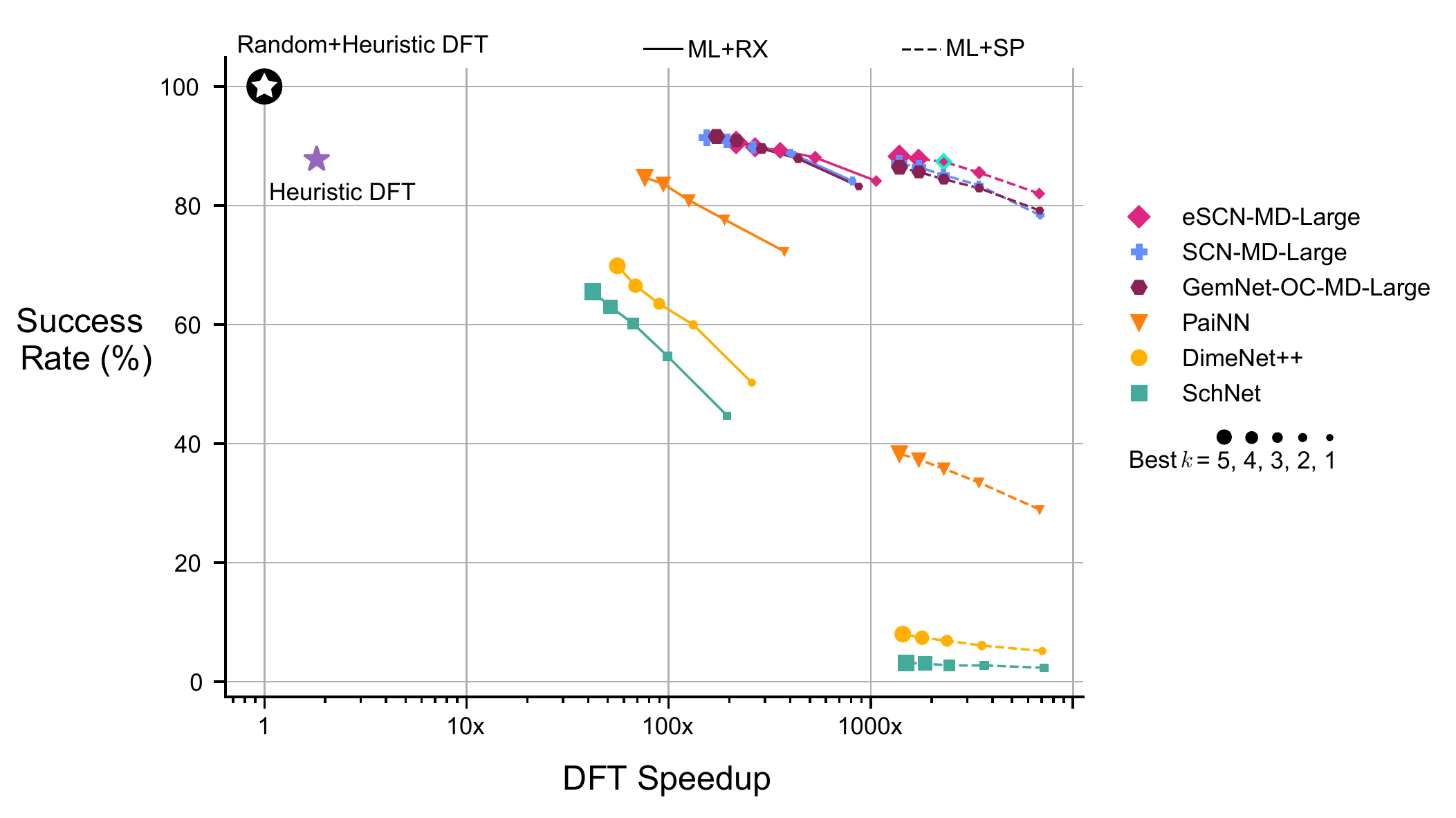}
    \caption{Overview of the accuracy-efficiency trade-offs of the proposed \algo{} methods across several baseline \gls{GNN} models. For each model, \gls{DFT} speedup and corresponding success rate are plotted for ML+RX and ML+SP across various best-$k$. A system is considered successful if the predicted adsorption energy is within 0.1 eV of the \gls{DFT} minimum, or lower. All success rates and speedups are relative to Random+Heuristic \gls{DFT}. Heuristic \gls{DFT} is shown as a common community baseline. The upper right-hand corner represent the optimal region - maximizing speedup and success rate. The point highlighted in teal corresponds to the balanced option reported in the abstract - a 87.36\% success rate and 2290x speedup. A similar figure for the \gls{OC20D} validation set can be found in the \gls{SI}.}
    \label{fig:test_success_v_speedup}
\end{figure*}

\subsection{Experiments}
Our goal is to find comparable or better adsorption energies to those found using \gls{DFT} alone in \gls{OC20D}. The metric we use to quantify this task is success rate, which is the percentage of \gls{OC20D} systems where our \gls{ML}+\gls{DFT} adsorption energy is within 0.1 eV or lower than the \gls{DFT} adsorption energy. A validation of the \gls{ML} energy is not included in these experiments since all final adsorption energies will come from at least a single \gls{DFT} call, ensuring all values are valid. Another metric we track is the speedup compared to the \gls{DFTHR} baseline. Speedup is evaluated as the ratio of \gls{DFT} electronic steps used by \gls{DFTHR} to the proposed hybrid \gls{ML}+\gls{DFT} strategy. A more detailed discussion on the metrics can be found in the Methods section. Unless otherwise noted, all results are reported on the test set, with results on the validation set found in the \gls{SI}. When evaluating the common baseline of \gls{DFTH} that uses only DFT calculations, a success rate of 87.76\% is achieved at a speedup of 1.81x.

\begin{figure*}[ht!]
    \centering
    \includegraphics[width=0.7\textwidth]{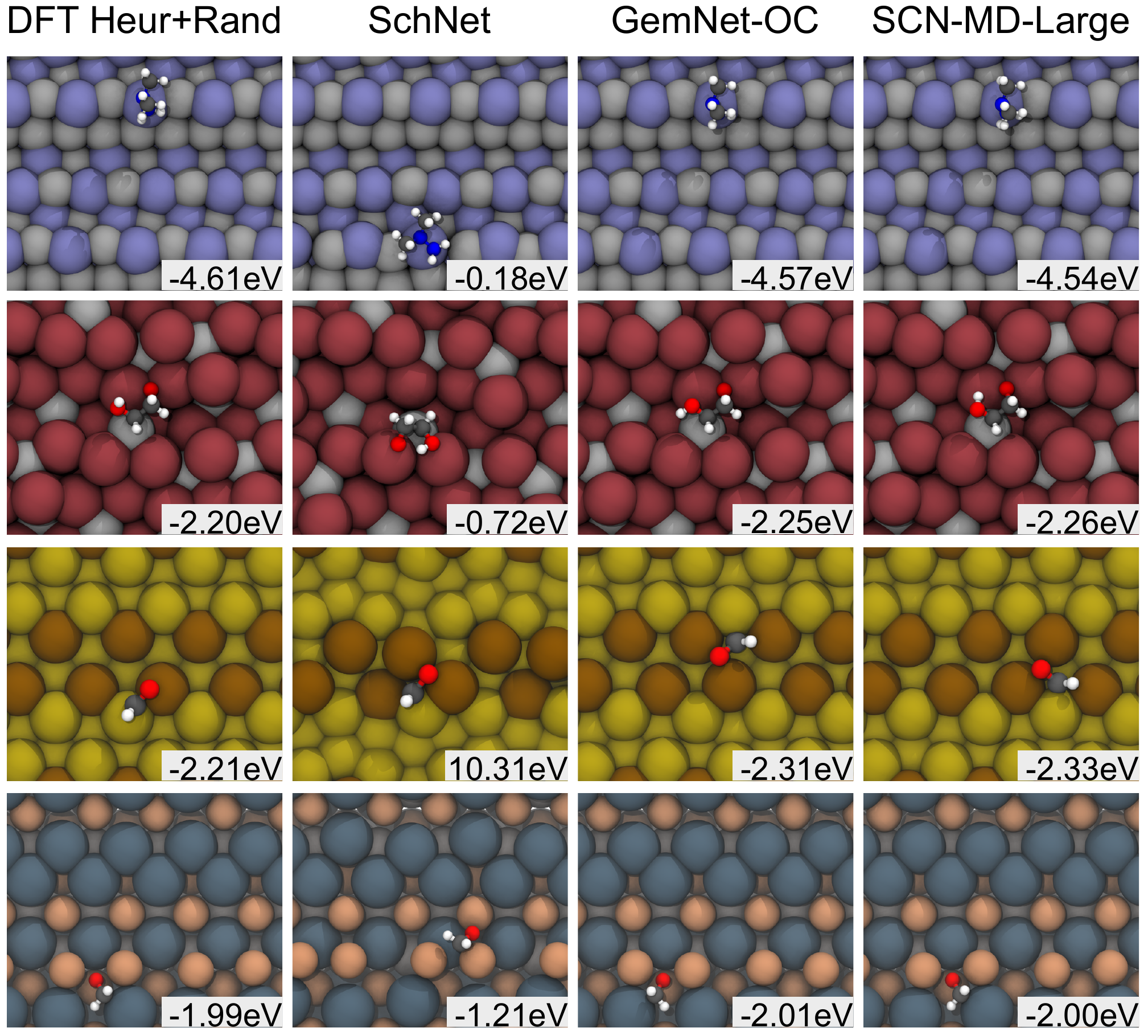}
    \caption{Illustration of the lowest energy configurations as found by \gls{DFTHR}, SchNet, GemNet-OC, and SCN-MD-Large on the \gls{OC20D} validation set. Corresponding adsorption energies are shown in the bottom right corner of each snapshot. \gls{ML} relaxed structures have energies calculated with a \gls{DFT} single-point, ML+SP. A variety of systems are shown including ones where \gls{ML} finds lower, higher, and comparable adsorption energies to \gls{DFT}. Notice that several of the configurations  in the third and fourth systems are symmetrically equivalent, and that SchNet induces a large surface reconstruction in the third system resulting in the extremely large DFT energy (10.31 eV).}
    \label{fig:placements}
\end{figure*}

\textbf{ML+SP} The results of using single-point evaluations on \gls{ML} relaxed states are summarized in Figure \ref{fig:test_success_v_speedup}. eSCN-MD-Large and GemNet-OC-MD-Large achieve a success rate of 86+\% at $k=5$ with eSCN-MD-Large outperforming all models with a success rate of 88.27\%, slightly better than the \gls{DFTH} baseline. Other models including SchNet and DimeNet++ do significantly worse with success metrics as low as 3.13\% and 7.99\%, respectively; suggesting the predicted relaxed structures are highly unfavorable. The speedups are fairly comparable across all models, ranging between 1400x and 1500x for k=5, orders of magnitude faster than the \gls{DFTH} baseline. Specifically, eSCN-MD-Large and GemNet-OC-MD-Large give rise to speedups of 1384x and 1388x, respectively. If speed is of most importance, speedups as high as 6817x are achievable with $k=1$ while still maintaining success rates of 82\% for eSCN-MD-Large. At a more balanced trade-off, $k=3$, success rates of 87.36\%  and 84.43\% are attainable for eSCN-MD-Large and GemNet-OC-MD-Large while maintaining speedups of 2296x and 2299x, respectively. In Figure \ref{fig:placements} the minimum energy binding sites of several systems are compared as identified with ML+SP across different models.

\input{tables/breakdown.tex}

\textbf{ML+RX} While single-point evaluations offer a fast evaluation of \gls{ML} structures, performance is heavily reliant on the accuracy of the predicted relaxed structure. This is particularly apparent when evaluating the max per-atom force norm of \gls{ML} relaxed structures with \gls{DFT}. SchNet and DimeNet++ have on average a max force, $f_{max}$, of 2.00 eV/\text{\AA} and 1.21eV/\text{\AA}, respectively, further supporting the challenge these models face in obtaining valid relaxed structures. On the other hand, models like GemNet-OC-MD-Large and eSCN-MD-Large have an average $f_{max}$ of 0.21eV/\text{\AA} and 0.15eV/\text{\AA}, respectively. While these models are a lot closer to valid relaxed structures (i.e. $f_{max}$ $\leq$ 0.05 eV/\text{\AA}), these results suggest that there is still room for further optimization. Results on \gls{DFT} relaxations from \gls{ML} relaxed states are plotted in Figure \ref{fig:test_success_v_speedup}. eSCN-MD-Large and GemNet-OC-MD-Large outperform all models at all $k$ values, with a 90.60\% and 91.61\% success rate at $k=5$, respectively. Given the additional \gls{DFT} costs associated with refining relaxations, speedups unsurprisingly decrease. At $k=5$, we see speedups of 215x and 172x for eSCN-MD-Large and GemNet-OC-MD-Large, respectively. Both SchNet and DimeNet++ see much smaller speedups at 42x and 55x, respectively. The much smaller speedups associated with SchNet and DimeNet++ suggest that a larger number of \gls{DFT} steps is necessary to relax the previously unfavorable configurations generated by the models. Conversely, eSCN-MD-Large's much larger speedup can be attributed to the near relaxed states (average $f_{max}\sim$0.15eV/\text{\AA}) it achieves in its predictions. With $k=1$, speedups of 1064x are achievable while still maintaining a success rate of 84.13\% for eSCN-MD-Large. At a more balanced trade-off, $k=3$, success rates of 89.28\%  and 89.59\% are attainable for eSCN-MD-Large and GemNet-OC-MD-Large while maintaining speedups of 356x and 288x, respectively. 

\begin{figure}[h!]
    \centering
    \includegraphics[width=0.4\textwidth]{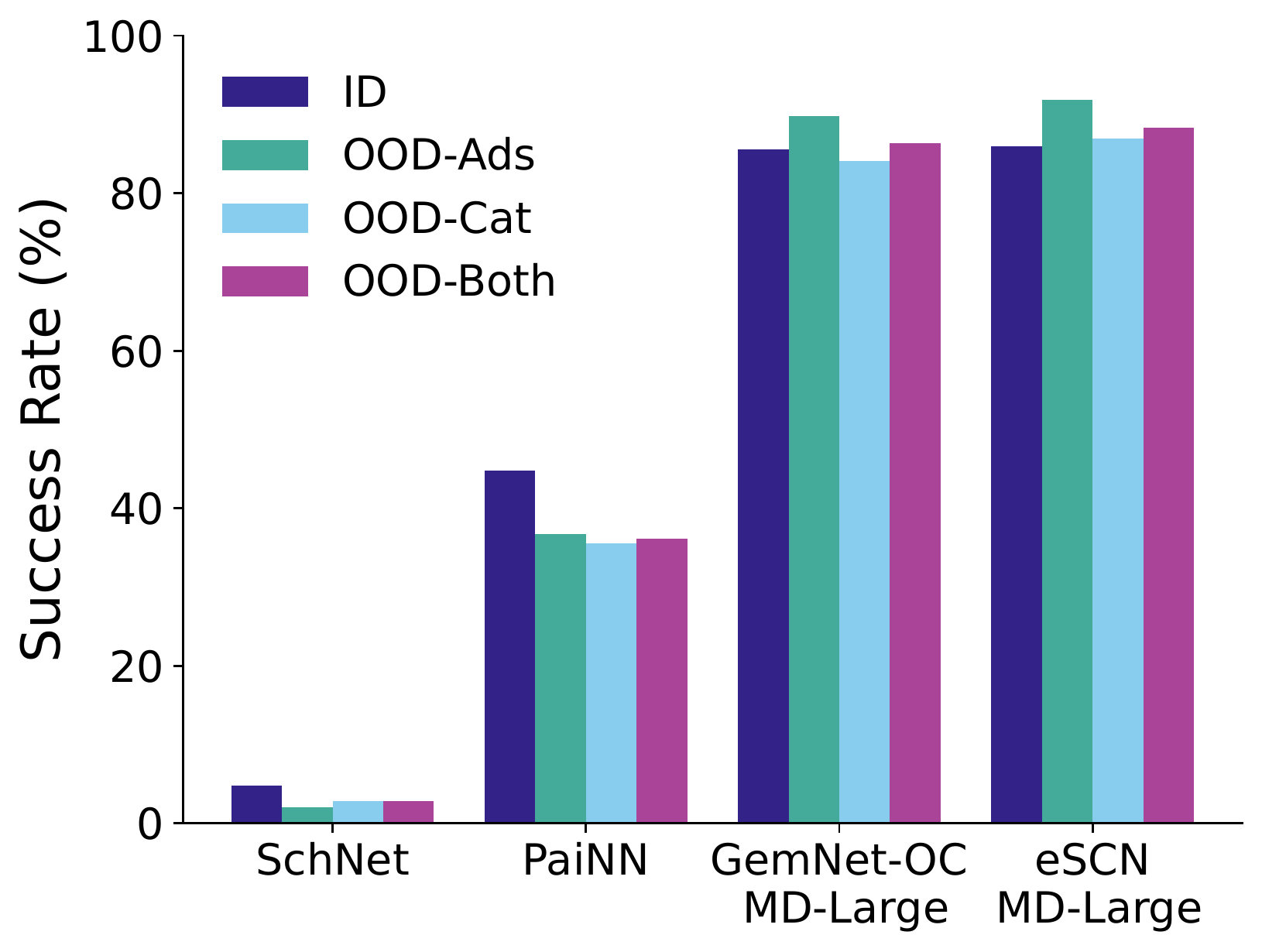}
    \caption{ML+SP success rate at $k=5$ across the different subsplits of the \gls{OC20D} test set and several baseline models. Top performing models show marginal differences across the different distribution splits, suggesting good generalization performance to out-of-domain adsorbates and catalysts not contained in the \gls{OC20} training dataset.}
    \label{fig:dist_bar}
\end{figure}

The results suggest a spectrum of accuracy and efficiency trade-offs that one should consider when selecting a strategy. For our best models, ML+SP results are almost 8x faster than ML+RX with only a marginal performance decrease in success rates (3-4\%), suggesting a worthwhile comprise. This difference is much more significant for worse models.

In Table \ref{tab:breakdowns} we measure the distribution of predictions that are much better, in parity, or much worse than the ground truth, where much better/worse corresponds to being lower/higher than 0.1 eV of the \gls{DFT} adsorption energy. Across both strategies, we observe that the most accurate models do not necessarily find much better minima. For instance, at $k=5$ ML+RX, eSCN-MD-Large finds 9.10\% of systems with much lower minima, compared to DimeNet++ finding 15.57\%. Similarly, while eSCN-MD-Large outperformed models in ML+SP, it observes less of an improvement with ML+RX; a consequence of the model arriving at a considerable local minima that a subsequent DFT relaxation has minimal benefit. This further suggests that some form of noise in models can aid in finding better minima. The full set of tabulated results for ML+SP and ML+RX experiments can be found in the \gls{SI} for the \gls{OC20D} test and validation sets.

\textbf{Distribution splits} Additionally, we evaluate success metrics across the different dataset subsplits. \gls{OC20D} uniformly samples from the four \gls{OC20} splits - \gls{ID}, \gls{OOD}-Adsorbate, \gls{OOD}-Catalyst, and \gls{OOD}-Both. Across our best models, we observe that performance remains consistent across the different distribution splits (Figure \ref{fig:dist_bar}). This suggests that for applications including adsorbates or surfaces that are not contained in OC20, AdsorbML still provides accurate and meaningful results. While we expect results to be consistent with \gls{OC20} where \gls{ID} outperforms \gls{OOD}, that is not necessarily the case here. eSCN-MD-Large, ML+SP at $k=5$, achieves 86.00\% on \gls{ID} while a 88.35\% success rate on \gls{OOD}-Both, with similar trends on ML+RX. We attribute this discrepancy to the fairly small sample size per split (250). The full set of results can be found in the \gls{SI}.

\textbf{Configuration analysis} Alongside the main results, we explore the performance of using only heuristic or only random \gls{ML} configurations on the \gls{OC20D} validation set. Results are reported on SCN-MD-Large, for the ML+SP strategy. At $k=5$, when only random configurations are used, success drops from 87.77\% to 82.94\%. More drastically, when only considering heuristic configurations, success drops significantly to 62.18\%. This suggests that random configurations can have a larger impact. Additional results can be found in the \gls{SI}.

%% file: tables/ml_success_analysis.tex
\begin{table*}[]
\begin{tabular}{@{}lcccc@{}}
\multicolumn{5}{c}{\gls{OC20D} Test} \\\toprule
\multirow{2}{*}{\textbf{Model}} & \multirow{2}{*}{\textbf{Success Rate* [\%] $\uparrow$}} & \multicolumn{1}{c|}{\multirow{2}{*}{\textbf{Energy MAE [eV] $\downarrow$}}} & \multicolumn{2}{c}{\textbf{OC20 S2EF MAE $\downarrow$}} \\
 &  & \multicolumn{1}{c|}{} & Forces [eV/\AA] & Energy [eV] \\ \midrule
SchNet & 1.01\% & \multicolumn{1}{c|}{0.5150} & 0.0496 & 0.4445 \\
DimeNet++ & 1.72\% & \multicolumn{1}{c|}{0.4329} & 0.0446 & 0.4753 \\
PaiNN & 10.92\% & \multicolumn{1}{c|}{0.2994} & 0.0294 & 0.2459 \\
GemNet-OC & 46.51\% & \multicolumn{1}{c|}{0.1849} & 0.0179 & 0.1668 \\
GemNet-OC-MD & 50.05\% & \multicolumn{1}{c|}{0.1966} & 0.0173 & 0.1694 \\
GemNet-OC-MD-Large & 48.03\% & \multicolumn{1}{c|}{0.1935} & 0.0164 & 0.1665 \\
SCN-MD-Large & 51.87\% & \multicolumn{1}{c|}{0.1758} & 0.0160 & 0.1730 \\ 
eSCN-MD-Large & 56.52\% & \multicolumn{1}{c|}{0.1739} & 0.0139 & 0.1709 \\ 
\midrule
\multicolumn{5}{l}{*ML predictions that lead to valid configurations and are within 0.1 eV of their DFT evaluation}
\end{tabular}
\caption{Success rates evaluated using \gls{ML} predicted energies. \gls{ML} predictions are only considered if their predicted energies are within 0.1 eV of its \gls{DFT} evaluation. Energy \gls{MAE} is also computed between predicted \gls{ML} and \gls{DFT} energy minima. We also show \gls{OC20} \gls{S2EF} Val-ID results, with metrics correlating well with success rates and energy \gls{MAE}.\label{tab:ml-success}}
\end{table*}

%% file: tables/splits.tex
\begin{table}[h!]
\resizebox{0.4\textwidth}{!}{%
\begin{tabular}{@{}lcccc@{}}
\toprule
\multirow{2}{*}{Split} & Unique & Unique & \multirow{2}{*}{Adsorbates} & \multirow{2}{*}{Bulks} \\
           & Systems & Configurations &    &     \\ \midrule
Validation & 973     & 85, 658        & 74 & 833 \\
Test       & 989     & 105,714        & 74 & 837 \\ \bottomrule
\end{tabular}%
}
\caption{Size of \gls{OC20D} validation and test splits. Unique adsorbate-surface systems are selected from the respective \gls{OC20} validation and test splits. Each split samples $\sim$250 systems from each of its respective distribution subsplits - \gls{ID}, \gls{OOD}-Ads, \gls{OOD}-Catalyst, \gls{OOD}-Both.}
\label{tab:oc20dense-splits}
\end{table}

%% file: tables/breakdown.tex
\begin{table*}[ht]
\resizebox{0.85\textwidth}{!}{%
\begin{tabular}{@{}lcccccc@{}}
\multicolumn{7}{c}{Success Rate} \\ \midrule
\multicolumn{7}{c}{DFT single-point on ML relaxed structures (ML+SP)} \\ \midrule
 & \multicolumn{3}{c|}{\textit{k=1}} & \multicolumn{3}{c}{\textit{k=5}} \\
\textbf{Model} & \multicolumn{1}{l}{Much better} & Parity & \multicolumn{1}{c|}{Much worse} & Much better & Parity & Much worse \\
SchNet & 0.40\% & 1.92\% & \multicolumn{1}{c|}{97.67\%} & 0.71\% & 2.43\% & 96.87\% \\
DimeNet++ & 0.91\% & 4.25\% & \multicolumn{1}{c|}{94.84\%} & 1.31\% & 6.67\% & 92.01\% \\
PaiNN & 2.12\% & 26.79\% & \multicolumn{1}{c|}{71.08\%} & 3.34\% & 34.98\% & 61.68\% \\
GemNet-OC & 6.47\% & 66.13\% & \multicolumn{1}{c|}{27.40\%} & 6.88\% & 74.12\% & 19.01\% \\
GemNet-OC-MD & 6.27\% & 70.17\% & \multicolumn{1}{c|}{23.56\%} & 7.58\% & 76.24\% & 16.18\% \\
GemNet-OC-MD-Large & 5.86\% & 73.31\% & \multicolumn{1}{c|}{20.83\%} & 7.18\% & 79.27\% & 13.55\% \\
SCN-MD-Large & 6.67\% & 71.69\% & \multicolumn{1}{c|}{21.64\%} & 7.58\% & 79.47\% & 12.94\% \\ 
eSCN-MD-Large & 5.06\% & 76.95\% & \multicolumn{1}{c|}{18.00\%} & 6.27\% & 82.00\% & 11.73\% \\ \midrule
\multicolumn{7}{c}{DFT relaxations on ML relaxed structures (ML+RX)} \\ \midrule
 & \multicolumn{3}{c|}{\textit{k=1}} & \multicolumn{3}{c}{\textit{k=5}} \\
\textbf{Model} & Much better & Parity & \multicolumn{1}{c|}{Much worse} & Much better & Parity & Much worse \\
SchNet & 10.82\% & 33.87\% & \multicolumn{1}{c|}{55.31\%} & 18.71\% & 46.81\% & 34.48\% \\
DimeNet++ & 9.40\% & 40.85\% & \multicolumn{1}{c|}{49.75\%} & 15.57\% & 54.30\% & 30.13\% \\
PaiNN & 9.81\% & 62.49\% & \multicolumn{1}{c|}{27.70\%} & 14.26\% & 70.48\% & 15.27\% \\
GemNet-OC & 9.81\% & 72.30\% & \multicolumn{1}{c|}{17.90\%} & 12.23\% & 75.73\% & 12.03\% \\
GemNet-OC-MD & 8.29\% & 74.12\% & \multicolumn{1}{c|}{17.59\%} & 11.63\% & 78.26\% & 10.11\% \\
GemNet-OC-MD-Large & 7.48\% & 75.73\% & \multicolumn{1}{c|}{16.78\%} & 10.11\% & 81.50\% & 8.39\% \\
SCN-MD-Large & 8.90\% & 75.23\% & \multicolumn{1}{c|}{15.87\%} & 12.94\% & 78.46\% & 8.59\% \\ 
eSCN-MD-Large & 6.47\% & 77.65\% & \multicolumn{1}{c|}{15.87\%} & 9.10\% & 81.50\% & 9.40\% \\ \bottomrule
\end{tabular}
}
\caption{\label{tab:breakdowns}Distribution of success rates for the proposed ML+SP and ML+RX strategies on the \gls{OC20D} test set. ``Parity" corresponds to being within 0.1 eV of the \gls{DFT} adsorption energy; ``Much better" corresponds to being less than 0.1 eV than \gls{DFT}; and ``Much worse" being higher than 0.1 eV of \gls{DFT}.}
\end{table*}

%% file: sections/discussion.tex
\section{Discussion}
We envision this work as an important but initial step towards reducing the computational cost of \gls{DFT} for not just catalysis applications, but computational chemistry more broadly. \algo{} provides a spectrum of accuracy and efficiency trade-offs one can choose depending on the application and computational resources available. For example, if we are interested in screening the largest number of \ch{CO2} reduction reaction catalysts possible, given a fixed compute budget, we could choose ML+SP at $k=2$ for a 85\% success rate while screening $\sim$3400x more materials than would have been possible with \gls{DFT} alone. On the other hand, if depth of study is more important, ML+RX is a good alternative as the structures are fully optimized with \gls{DFT} and the computational speedup comes from reducing the total number of relaxation steps required. In this scenario, the \gls{ML} potential serves as an efficient pre-optimization step. Even though \gls{ML} models comprise a small portion of the overall compute (see \gls{SI} for details), we expect these requirements to be reduced even further as more effort is placed on inference efficiency in the future.

One observation that merits additional studies is that \gls{ML} models found much better minima between 5\%-15\% of the time, depending on the efficiency trade-offs (Table~\ref{tab:breakdowns}). If our \gls{ML} models were perfect there would be no instances with lower adsorption energies; however, implicit noise in the form of inaccurate force predictions allows the \gls{ML} models to traverse unexplored regions of the potential energy surface. Exploring to what extent implicit and explicit noise~\cite{schaarschmidt2022learned, godwin2021simple} impact \gls{ML} relaxations and downstream tasks such as success rate is an important area of future research. 

Another natural extension to this work is focusing on alternative methods of global optimization and initial configuration generation. Here, we focused on accelerating brute force approaches to finding the global minimum by enumerating initial adsorbate-surface configurations. However, there are likely to be much more efficient approaches to global optimization such as minima hopping~\cite{goedecker2004minima}, constrained optimization~\cite{jung2022machine, Peterson2014}, Bayesian optimization, or a directly learned approach. It is worth noting that while our enumeration spanned a much larger space than traditional heuristic methods, it was not exhaustive and all-encompassing. We found that increasing the number of random configurations beyond what was sampled had diminishing returns, as the change in success rate from heuristic + 80\% random DFT to heuristic + 100\% random DFT was only 1.6\% (see the \gls{SI} for more details). If screening more \gls{ML} configurations continues to be advantageous, thinking about how we handle duplicate structures could further help accuracy and efficiency. We explore this briefly in the \gls{SI}, where removing systems with nearly the same \gls{ML} energies resulted in marginal benefit.

While current models like GemNet-OC and eSCN-MD-Large demonstrate impressive success rates on \gls{OC20D}, \gls{ML} relaxations without any subsequent \gls{DFT} are still not accurate enough for practical applications (Table \ref{tab:ml-success}). In order for future modeling work to address this challenge there are a number of observations worth highlighting. First, there is a positive correlation between success rate on \gls{OC20D} and both the \gls{S2EF} and relaxation based \gls{IS2RE} \gls{OC20} tasks. Thus, relaxation based \gls{IS2RE} and \gls{S2EF} metrics can be used as proxies when training models on \gls{OC20}. Another important note on model development is that \gls{OC20D}'s validation set is a subset of the \gls{OC20} validation set; as a result, the \gls{OC20} validation data should not be used for training when evaluating on \gls{OC20D}. Lastly, it is strongly encouraged that results reported on the \gls{OC20D} validation set be evaluated using a \gls{DFT} single-point calculation because the success rate metric can be manipulated by predicting only low energies. This could be done with as few as $\sim$1,000 single-point calculations. Alongside the release of the \gls{OC20D} test set, we will explore releasing a public evaluation server to ensure consistent evaluation and accessibility for \gls{DFT} evaluation, if there’s interest.

Tremendous progress in datasets and machine learning for chemistry has enabled models to reach the point where they can substantially enhance and augment \gls{DFT} calculations. Our results demonstrate that current state-of-the-art \gls{ML} models not only accelerate \gls{DFT} calculations for catalysis but enable more accurate estimates of properties that require global optimization such as adsorption energies. While the models used in this work are best suited for idealized adsorbate-surface catalysts, fine-tuning strategies can help enable applications to other chemistries including metal-organic frameworks and zeolites \cite{tran2022open}. Similarly, the models used in this work were trained on a consistent level of DFT theory (revised Perdew-Burke-Ernzerhof, no spin-polarization), generalizing to other functionals and levels of theory could also be enabled with fine-tuning or other training strategies. Given the timeline of \gls{ML} model development these results would not have been possible even a couple of years ago. We anticipate this work will accelerate the large-scale exploration of complex adsorbate-surface configurations for a broad range of chemistries and applications. Generalizing these results to more diverse materials and molecules without reliance on \gls{DFT} is a significant community challenge moving forward.

%% file: sections/methods.tex
\section{Methods} \label{methods} 

\subsection{\acrfull{OC20D}}

The evaluation of adsorption energy estimations requires a ground truth dataset that thoroughly explores the set of potential adsorption configurations. While \gls{OC20} computed adsorption energies for $O(1M)$ systems, the energies may not correspond to the minimum of that particular adsorbate-surface combination. More specifically, for a given catalyst surface, \gls{OC20} considers all possible adsorption sites but only places the desired adsorbate on a randomly selected site in one particular configuration. The tasks presented by \gls{OC20} enabled the development of more accurate machine learned potentials for catalysis ~\cite{gasteiger2022graph, zitnick2022spherical, ying2021transformers, godwin2021simple, shuaibi2021rotation}, but tasks like \gls{IS2RE}, although correlate well, are not always sufficient when evaluating performance as models are penalized when finding a different, lower energy minima - a more desirable outcome. As a natural extension to \gls{OC20}'s tasks, we introduce \gls{OC20D} to investigate the performance of models to finding the adsorption energy.

\gls{OC20D} is constructed to closely approximate the adsorption energy for a particular adsorbate-surface combination. To accomplish this, a dense sampling of initial adsorption configurations is necessary. \gls{OC20D} consists of two splits - a validation and test set. For each split, $\sim$1,000 unique adsorbate-surface combinations from the respective \gls{OC20} validation/test set are sampled. A uniform sample is then taken from each of the subsplits (\gls{ID}, \gls{OOD}-Adsorbate, \gls{OOD}-Catalyst, \gls{OOD}-Both) to explore the generalizability of models on this task. For each adsorbate-surface combination, two strategies were used to generate initial adsorbate configurations: heuristic and random. The heuristic strategy serves to represent the average catalysis researcher, where popular tools like CatKit~\cite{catkit} and Pymatgen~\cite{ong2013python} are used to make initial configurations. Given an adsorbate and surface, Pymatgen enumerates all symmetrically identical sites, an adsorbate is placed on the site, and a random rotation along the z axis 
 followed by slight wobbles in the x and y axis is applied to the adsorbate. While heuristic strategies seek to capture best practices, they do limit the possible search space with no guarantees that the true minimum energy is selected. To address this, we also randomly enumerate M sites on the surface and then place the adsorbate on top of the selected site. In this work, M=100 is used and a random rotation is applied to the adsorbate in a similar manner. In both strategies we remove unreasonable configurations - adsorbates not placed on the slab and/or placed too deep into the surface. \gls{DFT} relaxations were then run on all configurations with the results filtered to remove those that desorb, dissociate or create surface mismatches. The minimum energy across those remaining is considered the adsorption energy. While random is meant to be a more exhaustive enumeration, it is not perfect and could likely miss some adsorbate configurations. The \gls{OC20D} validation set was created in a similar manner but contained notable differences, details are outlined in the \gls{SI}.

The \gls{OC20D} test set comprises 989 unique adsorbate+surface combinations spanning 74 adsorbates and 837 bulks. Following the dense sampling, a total of 56,282 heuristic and 49,432 random configurations were calculated with \gls{DFT}. On average, there were 56 heuristic and 50 random configurations per system (note - while M=100 random sites were generated, less sites were available upon filtering.) In total, $\sim$4 million hours of compute were used to create the dataset. All \gls{DFT} calculations were performed using \gls{VASP}~\cite{Kresse1994, Kresse1996a, kresse1999ultrasoft, Kresse1996}. A discussion on \gls{DFT} settings and details can be found in the \gls{SI}.

\subsection{Evaluation Metrics}
To sufficiently track progress, we propose two primary metrics - success rate and \gls{DFT} speedup. \textbf{Success rate} is the proportion of systems in which a strategy returns energy that is within $\sigma$, or lower of the \gls{DFT} adsorption energy. A margin of $\sigma=0.1$eV is selected as the community is often willing to tolerate a small amount of error for practical relevance~\cite{kolluru2022open, Chanussot2021}. Tightening this threshold for improved accuracy is a foreseeable step once models+strategies saturate. While high success rates are achievable with increased \gls{DFT} compute, we use \textbf{\gls{DFT} speedup} as a means to evaluate efficiency. Speedup is measured as the ratio of \gls{DFT} electronic, or \gls{SC}, steps used by \gls{DFTHR} and the proposed strategy. Electronic steps are used as we have seen them correlate better with \gls{DFT} compute time than the number of ionic, or relaxation, steps. \gls{DFT} calculations that failed or resulted in invalid structures were included in speedup evaluation as they still represent realized costs in screening. We chose not to include compute time in this metric as results are often hardware dependent and can make comparing results unreliable. \gls{ML} relaxation costs are excluded from this metric as hardware variance along with CPU+GPU timings make it nontrivial to normalize. While \gls{ML} timings are typically negligible compared to the DFT calculations, a more detailed analysis of ML timings can be found in the \gls{SI}. Metrics are reported against the rigorous ground truth - \gls{DFTHR}, and compared to a community heuristic practice - \gls{DFTH}. Formally, metrics are defined in Equations  \ref{eq:success_rate} and \ref{eq:speedup}.

\begin{equation}\label{eq:success_rate}
\text{Success Rate}  =\frac{\sum_i^{N}\mathbbm{1}\big[\min(\hat{E}_{i}^{})-\min(E_{i}) \leq \sigma \big]}{N}
\end{equation}

\begin{equation}\label{eq:speedup}
\text{DFT Speedup} = \frac{\sum_{N} N_{SC steps}}{\sum_N \hat{N}_{SC steps}}
\end{equation}

\noindent where $i$ is an adsorbate-surface system, $N$ the total number of unique systems, $\mathbbm{1}(x)$ is the indicator function, $\hat{\square}$ is the proposed strategy, $N_{SC steps}$ is the number of self-consistency, or electronic steps, and $\min(E)$ is the minimum energy across all configurations of that particular system. For both metrics, higher is better.

\subsection{Relaxation Constraints}
It is possible that some of the adsorbate-surface configurations we consider may relax to a state that are necessary to discard in our analysis. For this work we considered three such scenarios: (1) desorption, (2) dissociation, and (3) significant adsorbate induced surface changes. Desorption, the adsorbate molecule not binding to the surface, is far less detrimental because desorbed systems are generally high energy. Still, it is useful to understand when none of the configurations considered have actually adsorbed to the surface. Dissociation, the breaking of an adsorbate molecule into different atoms or molecules, is problematic because the resulting adsorption energy is no longer consistent with what is of interest, i.e., the adsorption energy of a single molecule, not two or more smaller molecules. Including these systems can appear to correspond to lower adsorption energies, but due to the energy not representing the desired system it can result in false positives. Lastly, we also discard systems with significant adsorbate induced surface changes because, just as with dissociation, we are no longer calculating the energy of interest. In calculating adsorption energy, a term is included for the energy of the clean, relaxed surface. An underlying assumption in this calculation is that the corresponding adsorbate-surface system's resulting surface must be comparable to the corresponding clean surface, otherwise this referencing scheme fails and the resulting adsorption energy is inaccurate. For each of these instances we developed detection methods as a function of neighborhood connectivity, distance information, and atomic covalent radii. Depending on the user's application, one may decide to tighten the thresholds defined within. Details on each of the detection methods and further discussion can be found in the \gls{SI}.

%% file: sections/author_contributions.tex
A.P., L.Z., and Z.U. conceptualized the project and performed preliminary experiments. J.L., M.S., and B.M.W. substantially expanded the scope of the project, developed the final methodology, conducted all experiments, analyzed the results, and prepared the codebase and dataset for release under the guidance of Z.U. and L.Z. B.W. contributed to the methodology for detecting invalid configurations. A.D. contributed to the AdsorbML methodology and provided guidance on models and experiments. L.Z. and M.U. supervised the project. All authors contributed to the writing and editing of the paper. J.L, A.P., M.S., and B.M.W  contributed equally as co-first authors. 

%% file: sections/si_description.tex
\section{Supplementary Information}
The supplementary information contains all tabulated results, results figure for the \gls{OC20D} validation set, \gls{ML} model and compute details, \gls{OC20D} placement details, \gls{DFT} calculation details, details on the relaxation constraints, model constraint counts, unvalidated \gls{ML} success rates, and additional results on configuration analysis and random baselines.

%% file: sections/supplementary.tex
\section{\textbf{Supplementary Tables}}
\subsection{Main Paper Results}

Tabulated results are provided for both the \gls{OC20D} validation and test set. Supplementary Table~\ref{tab:main-dft-heur_rand} and Supplementary Table~\ref{tab:val-dft-heur_rand} evaluate against \acrlong{DFTHR}. Additionally, Supplementary Table~\ref{tab:main-dft-heur} and Supplementary Table~\ref{tab:val-dft-heur} evaluates against the less exhaustive, but more common \acrlong{DFTH} baseline.

\input{tables/main-dft-heur+rand.tex}
\input{tables/val-dft-heur+rand.tex}
\input{tables/main-dft-heur.tex}
\input{tables/val-dft-heur.tex}
\clearpage
\subsection{Subsplit Results}
Results evaluated across different subsplits are shown in Supplementary Table~\ref{tab:test-splits} and Supplementary Table~\ref{tab:val-splits}.
\input{tables/test_subsplits}
\input{tables/val_subsplits}
\clearpage

\subsection{Model Implementation and Compute Details}
Models used for this work included SchNet~\cite{schutt2017schnet}, DimeNett++~\cite{klicpera2020fast, klicpera2020directional}, PaiNN~\cite{schutt2021equivariant}, GemNet-OC~\cite{gasteiger2022graph}, GemNet-OC-MD~\cite{gasteiger2022graph}, GemNet-OC-MD-Large~\cite{gasteiger2022graph}, and SCN-MD-Large~\cite{zitnick2022spherical}. Note, while Gasteiger, et al.\cite{gasteiger2022graph} used two trained GemNet-OC-MD-Large models optimized for energy and forces to run relaxations and make \gls{IS2RE} predictions, we use only a single model, the force variant. No models were trained as part of this work, pretrained checkpoints were obtained directly from \modelsurl{} or by contacting the authors directly (SCN-MD-Large). All models used identical optimization parameters and ran for 300 relaxation steps or until that max per-atom force norm was less than or equal to 0.02 eV/\AA, whichever comes first. All model configuration files can be found at \configs{}.

While speedup metrics are defined solely based off \gls{DFT} electronic steps, the compute associated with \gls{ML} relaxations are reported in Supplementary Table~\ref{tab:compute} alongside the \gls{DFT} compute necessary for the example of evaluating the top $k=5$ systems. All model relaxations were done on 32GB NVIDIA V100 cards. 

\input{tables/compute.tex}
To consider both GPU and CPU timing we can compute an alternative speedup metric based off their total compute time:
\begin{align*}
\text{Alternative DFT Speedup} = \frac{\text{Total DFT Time}}{\text{Total ML+DFT Time}}
\end{align*}

To compare the impact of \gls{ML} compute time we consider the alternative speedup metric with and without factoring in \gls{ML} compute in the total time for the \gls{OC20D} validation set. Results are reported in Supplementary Table~\ref{tab:alt-speedup}. 
\input{tables/alt_speedup.tex} For larger model variants like SCN-MD-Large and GemNet-OC-MD-Large we see that \gls{ML} compute time is non-negligible, with speedups dropping from 3596x and 3885x to 1147x and 1686x, respectively when evaluating ML+SP at $k=1$. Smaller models like GemNet-OC, GemNet-OC-MD, and PaiNN see marginal drops in speedups. When considering ML+RX, the overall \gls{DFT} time involved in refining relaxations makes \gls{ML} compute a lot less significant, with the largest models like SCN-MD-Large and GemNet-OC-MD-Large seeing only a 24.6\% and 14.2\% slowdown. Also shown in Supplementary Table~\ref{tab:alt-speedup}, as $k$ is increased to 5, the compute associated with \gls{ML} becomes more insignificant. While \gls{ML} is often treated as negligible in workflows, it is important to be aware of the real cost, particularly when working at scale. These results suggest that strategies that leverage minimal \gls{DFT} (ML+SP) can often be bottlenecked by \gls{ML} compute if large, complex models are used like SCN-MD-Large. While leveraging the state-of-the-art model is often favorable, these results suggest that sacrificing a few percentage points on success rate could be a meaningful trade-off if we can increase throughput at inference (e.g. GemNet-OC vs SCN-MD-Large). We note that the models used in this work were used off the shelf, without optimizing for inference. There is significant potential to improve \gls{ML} throughput with adequate optimizations.

\subsection{Deduplication}
It is possible that different initial configurations relax to identical, or symmetrically identical sites with nearly identical \gls{ML} energies. As a result, this means that redundant \gls{DFT} calculations may be performed if such systems appear in the best $k$ ranking. Another way to look at this is that it is beneficial to have diverse candidates in the best $k$. This becomes more important if we increase the number of random placements.

One way to address this is through a deduplication step before selecting the best $k$ in the proposed algorithm. This would enable us to increase the number of random placements without the concern of redundant calculations. To explore this, we incorporate a deduplication step via \gls{DBSCAN}~\cite{schubert2017dbscan} to cluster configurations based off \gls{ML} relaxed energies. The best $k$ systems are then selected by looping through each cluster, taking the lowest energy of the group, and then removing it from the cluster until $k$ placements have been selected. Clusters are controlled by a hyperparameter $\Delta E$, specifying the maximum energy difference between points in a cluster. Too small of a $\Delta E$ can result in little deduplication while too large can result in unique systems being clustered together. Results on SCN-MD-Large ML+SP are reported in Supplementary Table~\ref{tab:dedup} for various $\Delta E$, with $\Delta E=0$ corresponding to no deduplication.

\input{tables/dedup.tex}

While we observe some improvements with deduplication, overall we see marginal benefit across all $k$. A $\Delta E$ of 0.01eV provides a minor improvement compared to no deduplication. More substantial improvements could come from exploring other strategies (e.g. structure-based) or increasing the number of placements. We leave these questions as potential future directions.

\subsection{Varying Heuristic+Random ratios}
While a fixed set of random configurations was generated for each system ($M=100$), an obvious question arises if more random configurations will aid in finding better minima. To explore whether a saturation point exists, we report results on \gls{DFTH} + varying proportion of random configurations in Supplementary Table~\ref{tab:heur+xrand}. While success rates unsurprisingly increase, we see diminishing returns with only a 1.6\% difference between 80\% and 100\% random configurations as compared to the 8\% improvement between 0\% and 10\% additional random configurations.

\input{tables/heur_xrand.tex}

\subsection{Additional Results}

\begin{figure}[h]
    \centering
    \includegraphics[width=\textwidth]{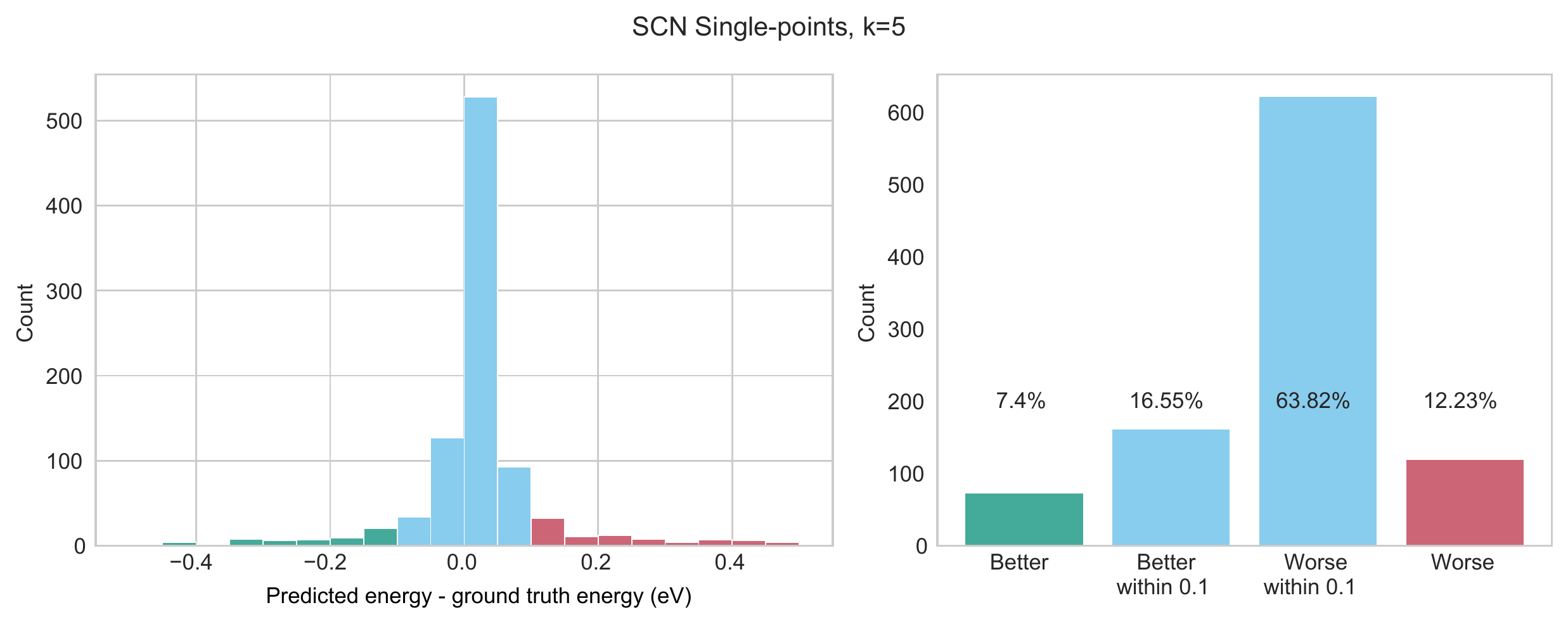}
    \includegraphics[width=\textwidth]{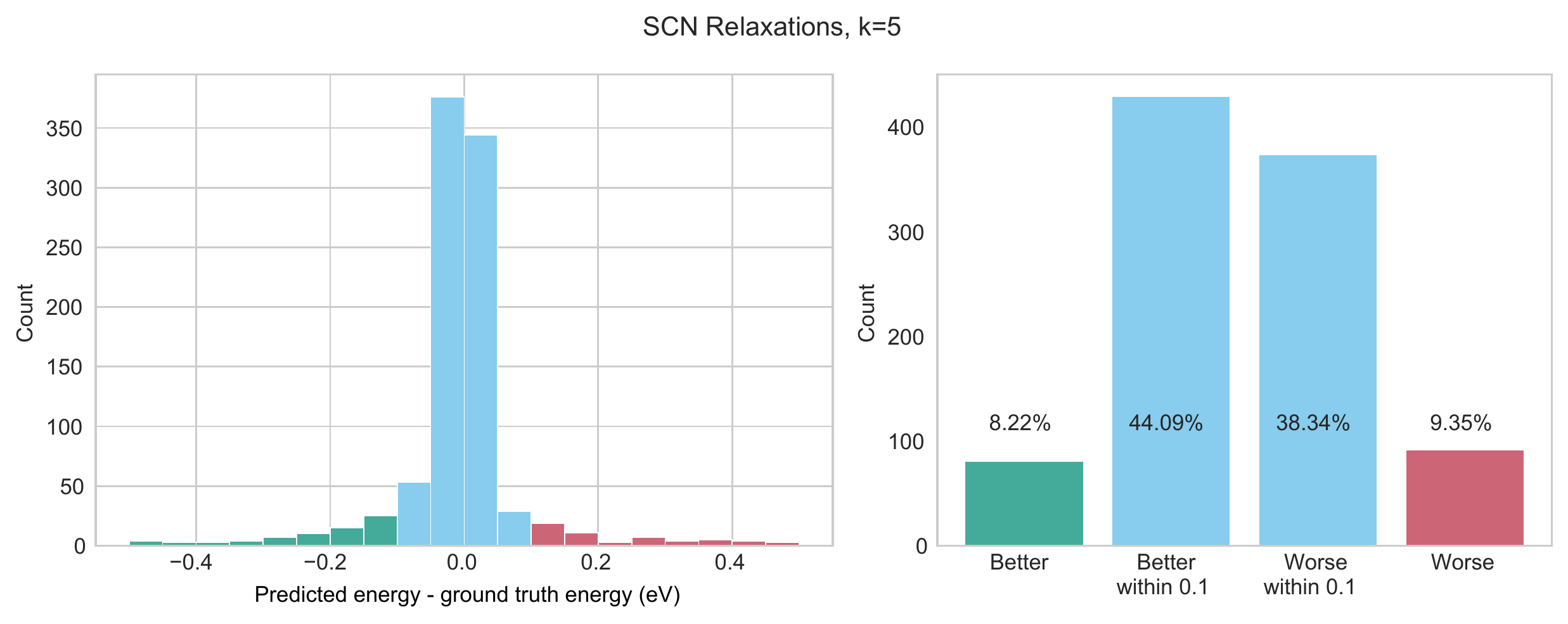}
    \caption{Results for SCN-MD-Large, single-points (top) and relaxations (bottom) at $k=5$. Left: distribution of differences between predicted and ground truth adsorption energies. Lower is better, meaning that \algo{} found a better binding site. Differences within 0.1 eV are also considered comparable and a success, represented in teal. Red bars are failure cases. Right: an aggregation of the major categories of energy differences. Results reported on the \gls{OC20D} validation set.}
    \label{fig:hist-scn}
\end{figure}

To better visualize the distribution of success rates, Supplementary Figure~\ref{fig:hist-scn} shows the breakdown for SCN-MD-Large. Even though the success rates of single-points and relaxations are similar, the more nuanced histogram shows how the predicted energies are lower with relaxations.

\subsubsection{Configuration analysis}

Supplementary Table~\ref{tab:comprandheur} compares the use of random and heuristic configurations independently. Random alone does slightly worse and heuristic alone does significantly worse when compared to the same ground truth. However, when limiting ground truth to the same set of initial configurations, success rates return to higher values.
\clearpage

\input{tables/compare-rand-heur.tex}
\subsubsection{Random baselines}

Supplementary Table~\ref{tab:rand-and-worst-k} shows success rates if we use ML to choose a different set of $k$ configurations, namely a random set and the worst set. These sanity checks confirm that the ML ranking of the best $k$ are indeed crucial, and that random and worst $k$ perform badly as expected.

\input{tables/random-and-worst}
\clearpage

\section{\textbf{Supplementary Figures}}
\subsection{OC20-Dense Validation Success v. Speedup}

\begin{figure*}[h]
    \centering
    \includegraphics[width=0.9\textwidth]{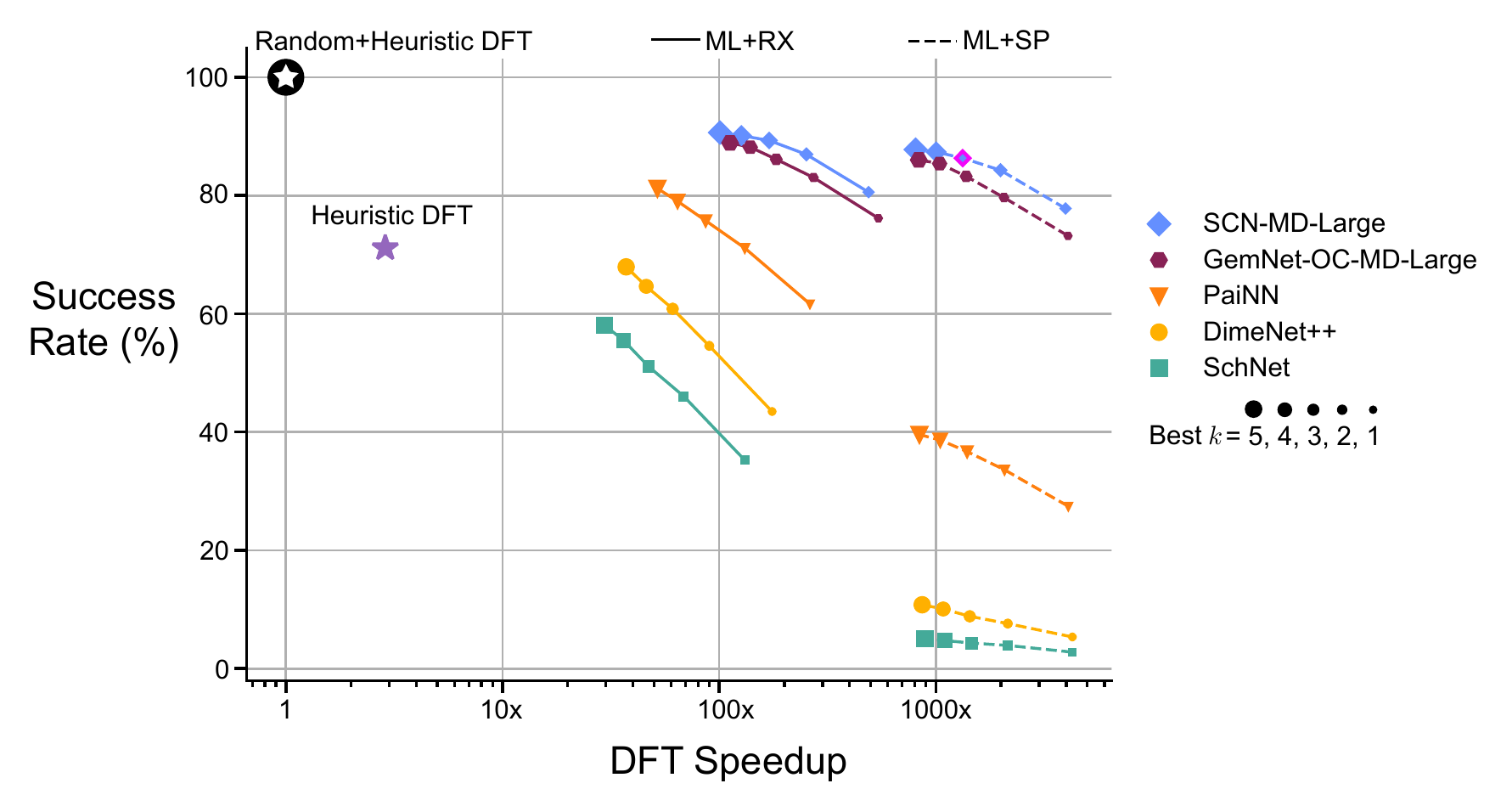}
    \caption{Overview of the accuracy-efficiency trade-offs of the proposed \algo{} methods across several baseline \gls{GNN} models on the \gls{OC20D} validation set. For each model, \gls{DFT} speedup and corresponding success rate are plotted for ML+RX and ML+SP across various best-$k$. A system is considered successful if the predicted adsorption energy is within 0.1 eV of the \gls{DFT} minimum, or lower. All success rates and speedups are relative to Random+Heuristic \gls{DFT}. Heuristic \gls{DFT} is shown as a common community baseline. The upper right-hand corner represent the optimal region - maximizing speedup and success rate. The point outlined in pink corresponds to the balanced option - a 86.33\% success rate and 1331x speedup.}
    \label{fig:success_v_speedup}
\end{figure*}

\section{\textbf{Supplementary Notes}}
\subsection{Relaxation Constraints}

To ensure proposed algorithms are accurately computing adsorption energies of the desired molecule, we filter problematic, or anomalous structures. These include dissociation, desorption, and adsorbate-induced surface changes. To accomplish this, we rely on neighborhood detection methods implemented in the \gls{ASE}~\cite{ase} detailed below.

To detect dissociation (1), a connectivity matrix is constructed for the adsorbate prior to relaxation and another is constructed for the adsorbate after relaxation. Two atoms are considered connected if the covalent radii have any overlap. The two matrices are compared and must be identical, otherwise it is classified as dissociated. To detect desorption, a connectivity matrix is constructed for the relaxed adsorbate-surface configuration. In this case, atoms are considered connected if there is any overlap of the atomic radii with a small cushion. This cushion is a 1.5 multiplier to the covalent radii. We did this so that we would only discard systems where the adsorbate has no interaction with the surface to avoid discarding physisorbed systems. To detect significant adsorbate induced surface changes (3), a connectivity matrix is constructed for the relaxed surface and another is constructed for the relaxed surface-adsorbate configuration. For the surface-adsorbate configuration, the subset of atoms belonging to the surface are considered. The process of constructing the connectivity matrices is repeated twice. First, with a cushion applied to the relaxed surface but no cushion applied to the relaxed adsorbate-surface configuration. Second, with a cushion applied to the relaxed adsorbate-surface configuration but no cushion applied to the relaxed surface. This cushion is a 1.5 multiplier to the covalent radii. For each of these cases, we check that the connected atoms for the system without the cushion is a subset of those found with the cushion. Considering both cases ensures that we are considering both bond breaking and bond forming events and are not ignoring cases where bonds are only broken as would occur if a surface atom moved up into the vacuum layer.

\subsection{Constraint Counts}
For all models, \gls{ML} relaxations were removed that violated certain physical constraints(dissociation, desorption, surface mismatch). Supplementary Table~\ref{tab:anomalies} shares a breakdown of the filtered counts for different models on the \gls{OC20D} validation set. Unsurprisingly, top performing models like SCN-MD-Large and GemNet-OC have a lot fewer removed and more comparable to \gls{DFT} than models like SchNet and DimeNet++. 

\input{tables/anomalies.tex}
\subsection{DFT and Calculation Details}
\gls{DFT} relaxations were performed consistent with \gls{OC20}'s methodology. \acrfull{VASP} with \gls{PAW} pseudopotentials and the \gls{RPBE} functional were used for all calculations~\cite{Kresse1994, Kresse1996a, kresse1999ultrasoft, Kresse1996}. All relaxations were performed with a maximum number of electronic steps of 60. All single-point evaluations were allowed a maximum of 300 electronic steps. This was done to ensure that the initialized wavefunction had sufficient steps to converge. Single-point calculations in which electronic steps were unconverged were discarded. The same was done for unconverged electronic steps at relaxed structures for relaxation calculations. All other settings and details are consistent with the \gls{OC20} manuscript~\cite{Chanussot2021}.

Similarly, adsorption energy calculations are also done consistent with \gls{OC20}. We note that there is some ambiguity in the catalysis literature for the choice of the gas phase reference, $E_{gas}$. If the adsorbate is itself a stable gas phase molecule then the adsorption energy might be calculated referenced to itself in the gas phase. However, this quantity is less helpful when calculating thermodynamically consistent free energy diagrams. As used in this work,  $E_{gas}$ is often chosen as a linear combination of reference gas phase species ~\cite{Chanussot2021, garcia2019statistical, gao2020determining}.

\subsection{OC20-Dense Placement Details}

For a unique adsorbate-surface combination, multiple adsorbate configurations were enumerated as part of the proposed \algo{} pipeline. The \gls{OC20D} validation and test set were created at different stages of the manuscript, with some notable placement code improvements happening between the two. We highlight those changes here.

The \gls{OC20D} validation set was created using the code provided at \url{https://github.com/Open-Catalyst-Project/Open-Catalyst-Dataset/tree/86b5254fe5}. There, the heuristic strategy used CatKit\cite{catkit} to enumerate all symmetrically distinct sites and provide a suggested adsorbate orientation. The random strategy randomly enumerated
M=100 configurations on the surface and placed the adsorbate 2 \r{A} above (in the z direction) the selected site. A random rotation is then applied to the adsorbate along the (0,0,1) adsorption site vector. The \gls{OC20D} test set was created slightly different following improvements to code provided at \url{https://github.com/Open-Catalyst-Project/Open-Catalyst-Dataset/tree/628c5136d0}. For random, sites are defined by first constructiung a Delaunay meshgrid with surface atoms as nodes. The positions of the sites are uniformly randomly sampled along the Delaunay triangles. For heuristic, we use the functionality built in Pymatgen\cite{ong2013python}, which similarly makes a Delaunay meshgrid. We consider sites on the nodes (atop), between 2 nodes (bridge) and in the centers of the triangles (hollow). For both approaches, the adsorbate is uniformly randomly rotated around the z direction, and provided a slight wobble around x and y, which amounts to randomized tilt within a certain cone around the north pole. The adsorbate database includes information about which atoms are expected to bind. The binding atom of the adsorbate is placed at the site. After being placed at the site, the adsorbate is translated along the surface normal until it is no longer overlapping with the surface and the minimum distance between any adsorbate and surface atom is 0.1 \r{A}. Despite the differences, results across all models between the two splits retain the same trends, supporting the use of the validation set for development. The improved heuristic strategy is reflected in the difference in the \gls{DFTH} baseline - 87.76\% and 1.81x speedup vs. 71.12\% and 2.87x for the \gls{OC20D} test and validation set, respectively.

%% file: tables/main-dft-heur+rand.tex
\begin{table*}[h]
\resizebox{0.75\textwidth}{!}{%
\begin{tabular}{lcccccccccc}
\multicolumn{11}{c}{OC20-Dense Test} \\ \hline
\multicolumn{11}{c}{\textbf{ML+DFT Singlepoints (ML+SP)}} \\ \hline
\textbf{Model} &
  \multicolumn{2}{c}{\textit{\textbf{k=1}}} &
  \multicolumn{2}{c}{\textit{\textbf{k=2}}} &
  \multicolumn{2}{c}{\textit{\textbf{k=3}}} &
  \multicolumn{2}{c}{\textit{\textbf{k=4}}} &
  \multicolumn{2}{c}{\textit{\textbf{k=5}}} \\
\textbf{} &
  \textbf{Success} &
  \textbf{Speedup} &
  \textbf{Success} &
  \textbf{Speedup} &
  \textbf{Success} &
  \textbf{Speedup} &
  \textbf{Success} &
  \textbf{Speedup} &
  \textbf{Success} &
  \textbf{Speedup} \\
SchNet &
  2.33\% &
  7202.16 &
  2.73\% &
  3645.62 &
  2.73\% &
  2446.02 &
  3.03\% &
  1854.64 &
  3.13\% &
  1496.95 \\
DimeNet++ &
  5.16\% &
  7057.96 &
  6.07\% &
  3544.06 &
  6.88\% &
  2380.24 &
  7.38\% &
  1794.72 &
  7.99\% &
  1439.84 \\
PaiNN &
  28.92\% &
  6841.82 &
  33.47\% &
  3428.48 &
  35.79\% &
  2297.60 &
  37.31\% &
  1727.91 &
  38.32\% &
  1385.31 \\
GemNet-OC &
  72.60\% &
  6828.17 &
  76.85\% &
  3428.86 &
  79.27\% &
  2302.96 &
  80.18\% &
  1732.48 &
  80.99\% &
  1389.94 \\
GemNet-OC-MD &
  76.44\% &
  6803.10 &
  80.79\% &
  3425.00 &
  82.61\% &
  2291.70 &
  83.32\% &
  1724.68 &
  83.82\% &
  1383.52 \\
GemNet-OC-MD-Large &
  79.17\% &
  6856.28 &
  82.91\% &
  3440.35 &
  84.43\% &
  2299.03 &
  85.64\% &
  1729.76 &
  86.45\% &
  1388.03 \\
SCN-MD-Large &
  78.36\% &
  6878.87 &
  83.42\% &
  3420.15 &
  85.14\% &
  2290.25 &
  86.45\% &
  1725.47 &
  87.06\% &
  1383.26 \\ 
eSCN-MD-Large &
  82.00\% &
  6817.21 &
  85.54\% &
  3437.52 &
  87.36\% &
  2296.16 &
  87.87\% &
  1724.63 &
  88.27\% &
  1384.10 \\ \hline
\multicolumn{11}{c}{\textbf{ML+DFT Relaxations (ML+RX)}} \\ \hline
\textbf{Model} &
  \multicolumn{2}{c}{\textit{\textbf{k=1}}} &
  \multicolumn{2}{c}{\textit{\textbf{k=2}}} &
  \multicolumn{2}{c}{\textit{\textbf{k=3}}} &
  \multicolumn{2}{c}{\textit{\textbf{k=4}}} &
  \multicolumn{2}{c}{\textit{\textbf{k=5}}} \\
\textbf{} &
  \textbf{Success} &
  \textbf{Speedup} &
  \textbf{Success} &
  \textbf{Speedup} &
  \textbf{Success} &
  \textbf{Speedup} &
  \textbf{Success} &
  \textbf{Speedup} &
  \textbf{Success} &
  \textbf{Speedup} \\
SchNet &
  44.69\% &
  194.66 &
  54.70\% &
  98.72 &
  60.16\% &
  66.70 &
  62.99\% &
  51.61 &
  65.52\% &
  42.08 \\
DimeNet++ &
  50.25\% &
  257.78 &
  59.96\% &
  132.39 &
  63.50\% &
  89.82 &
  66.53\% &
  68.52 &
  69.87\% &
  55.73 \\
PaiNN &
  72.30\% &
  373.54 &
  77.65\% &
  189.02 &
  80.89\% &
  126.04 &
  83.62\% &
  94.07 &
  84.73\% &
  76.26 \\
GemNet-OC &
  82.10\% &
  727.68 &
  85.64\% &
  372.27 &
  87.06\% &
  252.36 &
  87.06\% &
  190.00 &
  87.97\% &
  151.89 \\
GemNet-OC-MD &
  82.41\% &
  759.83 &
  86.55\% &
  392.89 &
  88.27\% &
  260.95 &
  89.18\% &
  193.01 &
  89.89\% &
  154.99 \\
GemNet-OC-MD-Large &
  83.22\% &
  872.63 &
  87.87\% &
  437.01 &
  89.59\% &
  288.80 &
  90.90\% &
  216.86 &
  91.61\% &
  172.46 \\
SCN-MD-Large &
  84.13\% &
  811.03 &
  88.78\% &
  403.70 &
  89.79\% &
  262.88 &
  90.90\% &
  194.10 &
  91.41\% &
  154.31 \\
eSCN-MD-Large &
  84.13\% &
  1064.42 &
  88.07\% &
  530.80 &
  89.28\% &
  356.25 &
  89.79\% &
  267.78 &
  90.60\% &
  215.58 \\ \bottomrule
\end{tabular}
}
\caption{Model success and speedup results as evaluated against \gls{DFT}-Heuristic+Random across varying $k$ for the OC20-Dense test set. This evaluation corresponds to a more exhaustive, but expensive approach - reflected by the increased speedups.\label{tab:main-dft-heur_rand}}
\end{table*}

%% file: tables/val-dft-heur+rand.tex
\begin{table*}[h]
\resizebox{0.75\textwidth}{!}{%
\begin{tabular}{@{}lcccccccccc@{}}
\multicolumn{11}{c}{OC20-Dense Validation} \\ \hline
\multicolumn{11}{c}{\textbf{ML+DFT Single-points (ML+SP)}} \\ \midrule
\textbf{Model} &
  \multicolumn{2}{c}{\textit{\textbf{k=1}}} &
  \multicolumn{2}{c}{\textit{\textbf{k=2}}} &
  \multicolumn{2}{c}{\textit{\textbf{k=3}}} &
  \multicolumn{2}{c}{\textit{\textbf{k=4}}} &
  \multicolumn{2}{c}{\textit{\textbf{k=5}}} \\
\textbf{} &
  \textbf{Success} &
  \textbf{Speedup} &
  \textbf{Success} &
  \textbf{Speedup} &
  \textbf{Success} &
  \textbf{Speedup} &
  \textbf{Success} &
  \textbf{Speedup} &
  \textbf{Success} &
  \textbf{Speedup} \\
SchNet &
  2.77\% &
  4266.13 &
  3.91\% &
  2155.36 &
  4.32\% &
  1458.77 &
  4.73\% &
  1104.88 &
  5.04\% &
  892.79 \\
DimeNet++ &
  5.34\% &
  4271.23 &
  7.61\% &
  2149.78 &
  8.84\% &
  1435.21 &
  10.07\% &
  1081.96 &
  10.79\% &
  865.20 \\
PaiNN &
  27.44\% &
  4089.77 &
  33.61\% &
  2077.65 &
  36.69\% &
  1395.55 &
  38.64\% &
  1048.63 &
  39.57\% &
  840.44 \\
GemNet-OC &
  68.76\% &
  4185.18 &
  77.29\% &
  2087.11 &
  80.78\% &
  1392.51 &
  81.50\% &
  1046.85 &
  82.94\% &
  840.25 \\
GemNet-OC-MD &
  68.76\% &
  4182.04 &
  78.21\% &
  2092.27 &
  81.81\% &
  1404.11 &
  83.25\% &
  1053.36 &
  84.38\% &
  841.64 \\
GemNet-OC-MD-Large &
  73.18\% &
  4078.76 &
  79.65\% &
  2065.15 &
  83.25\% &
  1381.39 &
  85.41\% &
  1041.50 &
  86.02\% &
  834.46 \\
SCN-MD-Large &
  77.80\% &
  3974.21 &
  84.28\% &
  1989.32 &
  86.33\% &
  1331.43 &
  87.36\% &
  1004.40 &
  87.77\% &
  807.00 \\ \midrule
\multicolumn{11}{c}{\textbf{ML+DFT Relaxations (ML+RX)}} \\ \midrule
\textbf{Model} &
  \multicolumn{2}{c}{\textit{\textbf{k=1}}} &
  \multicolumn{2}{c}{\textit{\textbf{k=2}}} &
  \multicolumn{2}{c}{\textit{\textbf{k=3}}} &
  \multicolumn{2}{c}{\textit{\textbf{k=4}}} &
  \multicolumn{2}{c}{\textit{\textbf{k=5}}} \\
\textbf{} &
  \textbf{Success} &
  \textbf{Speedup} &
  \textbf{Success} &
  \textbf{Speedup} &
  \textbf{Success} &
  \textbf{Speedup} &
  \textbf{Success} &
  \textbf{Speedup} &
  \textbf{Success} &
  \textbf{Speedup} \\
SchNet &
  35.25\% &
  131.26 &
  46.04\% &
  68.64 &
  51.08\% &
  47.24 &
  55.50\% &
  36.19 &
  58.07\% &
  29.58 \\
DimeNet++ &
  43.47\% &
  175.54 &
  54.57\% &
  90.04 &
  60.84\% &
  61.01 &
  64.65\% &
  46.06 &
  67.93\% &
  37.21 \\
PaiNN &
  61.66\% &
  262.38 &
  71.12\% &
  131.30 &
  75.75\% &
  86.64 &
  79.03\% &
  64.27 &
  81.19\% &
  51.88 \\
GemNet-OC &
  73.59\% &
  448.50 &
  83.14\% &
  231.76 &
  86.84\% &
  152.57 &
  88.18\% &
  117.40 &
  89.41\% &
  95.24 \\
GemNet-OC-MD &
  72.25\% &
  503.62 &
  81.40\% &
  251.69 &
  85.10\% &
  167.71 &
  87.05\% &
  124.21 &
  88.49\% &
  100.21 \\
GemNet-OC-MD-Large &
  76.16\% &
  543.48 &
  83.04\% &
  272.66 &
  86.13\% &
  183.47 &
  88.18\% &
  139.20 &
  88.90\% &
  112.29 \\
SCN-MD-Large &
  80.58\% &
  489.29 &
  86.95\% &
  252.77 &
  89.31\% &
  170.15 &
  90.13\% &
  126.72 &
  90.65\% &
  100.92 \\ \bottomrule
\end{tabular}%
}
\caption{Model success and speedup results as evaluated against \gls{DFT}-Heuristic+Random across varying $k$ for the OC20-Dense validation set. This evaluation corresponds to a more exhaustive, but expensive approach - reflected by the increased speedups.\label{tab:val-dft-heur_rand}}
\end{table*}

%% file: tables/main-dft-heur.tex
\begin{table*}[h]
\resizebox{0.80\textwidth}{!}{%
\begin{tabular}{lcccccccccc}
\multicolumn{11}{c}{OC20-Dense Test} \\ \hline
\multicolumn{11}{c}{\textbf{ML+DFT Singlepoints (ML+SP)}} \\ \hline
\textbf{Model} &
  \multicolumn{2}{c}{\textit{\textbf{k=1}}} &
  \multicolumn{2}{c}{\textit{\textbf{k=2}}} &
  \multicolumn{2}{c}{\textit{\textbf{k=3}}} &
  \multicolumn{2}{c}{\textit{\textbf{k=4}}} &
  \multicolumn{2}{c}{\textit{\textbf{k=5}}} \\
\textbf{} &
  \textbf{Success} &
  \textbf{Speedup} &
  \textbf{Success} &
  \textbf{Speedup} &
  \textbf{Success} &
  \textbf{Speedup} &
  \textbf{Success} &
  \textbf{Speedup} &
  \textbf{Success} &
  \textbf{Speedup} \\
SchNet &
  2.88\% &
  4006.55 &
  3.40\% &
  2027.43 &
  3.50\% &
  1359.96 &
  3.81\% &
  1030.36 &
  4.01\% &
  831.24 \\
DimeNet++ &
  6.48\% &
  3917.46 &
  7.61\% &
  1964.28 &
  8.64\% &
  1318.71 &
  9.05\% &
  993.91 &
  9.47\% &
  797.17 \\
PaiNN &
  33.64\% &
  3817.50 &
  38.07\% &
  1909.48 &
  40.43\% &
  1278.98 &
  41.87\% &
  961.09 &
  42.90\% &
  770.42 \\
GemNet-OC &
  76.85\% &
  3799.78 &
  81.28\% &
  1906.74 &
  83.54\% &
  1279.51 &
  84.36\% &
  961.73 &
  84.88\% &
  771.19 \\
GemNet-OC-MD &
  79.84\% &
  3792.77 &
  83.74\% &
  1909.18 &
  85.70\% &
  1275.16 &
  86.73\% &
  958.61 &
  87.04\% &
  768.24 \\
GemNet-OC-MD-Large &
  82.82\% &
  3816.00 &
  86.11\% &
  1914.08 &
  87.65\% &
  1278.30 &
  88.48\% &
  961.07 &
  89.40\% &
  770.73 \\
SCN-MD-Large &
  82.61\% &
  3838.32 &
  87.24\% &
  1905.32 &
  88.79\% &
  1275.08 &
  89.92\% &
  959.76 &
  90.33\% &
  768.69 \\
eSCN-MD-Large &
  85.60\% &
  3795.10 &
  88.79\% &
  1913.05 &
  90.53\% &
  1277.51 &
  90.95\% &
  958.69 &
  91.46\% &
  768.80 \\ \hline
\multicolumn{11}{c}{\textbf{ML+DFT Relaxations (ML+RX)}} \\ \hline
\textbf{Model} &
  \multicolumn{2}{c}{\textit{\textbf{k=1}}} &
  \multicolumn{2}{c}{\textit{\textbf{k=2}}} &
  \multicolumn{2}{c}{\textit{\textbf{k=3}}} &
  \multicolumn{2}{c}{\textit{\textbf{k=4}}} &
  \multicolumn{2}{c}{\textit{\textbf{k=5}}} \\
\textbf{} &
  \textbf{Success} &
  \textbf{Speedup} &
  \textbf{Success} &
  \textbf{Speedup} &
  \textbf{Success} &
  \textbf{Speedup} &
  \textbf{Success} &
  \textbf{Speedup} &
  \textbf{Success} &
  \textbf{Speedup} \\
SchNet &
  47.43\% &
  108.43 &
  57.92\% &
  55.02 &
  63.17\% &
  37.14 &
  66.15\% &
  28.72 &
  68.42\% &
  23.43 \\
DimeNet++ &
  54.22\% &
  143.15 &
  63.99\% &
  73.61 &
  67.80\% &
  49.86 &
  70.78\% &
  38.00 &
  74.07\% &
  30.90 \\
PaiNN &
  76.54\% &
  208.76 &
  81.89\% &
  105.98 &
  84.88\% &
  70.64 &
  87.35\% &
  52.68 &
  88.68\% &
  42.66 \\
GemNet-OC &
  85.91\% &
  414.29 &
  89.40\% &
  209.98 &
  90.84\% &
  141.75 &
  90.95\% &
  106.41 &
  91.67\% &
  85.01 \\
GemNet-OC-MD &
  86.01\% &
  429.26 &
  89.40\% &
  222.61 &
  91.05\% &
  148.26 &
  91.67\% &
  109.18 &
  92.08\% &
  87.38 \\
GemNet-OC-MD-Large &
  86.21\% &
  502.76 &
  90.53\% &
  248.12 &
  92.08\% &
  163.33 &
  93.11\% &
  122.23 &
  93.52\% &
  96.86 \\
SCN-MD-Large &
  87.45\% &
  454.12 &
  91.98\% &
  226.07 &
  92.80\% &
  147.38 &
  93.62\% &
  108.85 &
  94.03\% &
  86.29 \\
eSCN-MD-Large &
  87.55\% &
  594.32 &
  91.15\% &
  296.49 &
  92.28\% &
  198.63 &
  92.80\% &
  149.40 &
  93.62\% &
  120.13 \\ \bottomrule
\end{tabular}
}
\caption{Model success and speedup results as evaluated against \gls{DFT}-Heuristic across varying $k$ for the OC20-Dense test set. This evaluation corresponds to a more common community approach.\label{tab:main-dft-heur}}
\end{table*}

%% file: tables/val-dft-heur.tex
\begin{table*}[h]
\resizebox{0.80\textwidth}{!}{%
\begin{tabular}{@{}lcccccccccc@{}}
\multicolumn{11}{c}{OC20-Dense Validation} \\ \hline
\multicolumn{11}{c}{\textbf{ML+DFT Single-points (ML+SP)}} \\ \midrule
\textbf{Model} &
  \multicolumn{2}{c}{\textit{\textbf{k=1}}} &
  \multicolumn{2}{c}{\textit{\textbf{k=2}}} &
  \multicolumn{2}{c}{\textit{\textbf{k=3}}} &
  \multicolumn{2}{c}{\textit{\textbf{k=4}}} &
  \multicolumn{2}{c}{\textit{\textbf{k=5}}} \\
\textbf{} &
  \textbf{Success} &
  \textbf{Speedup} &
  \textbf{Success} &
  \textbf{Speedup} &
  \textbf{Success} &
  \textbf{Speedup} &
  \textbf{Success} &
  \textbf{Speedup} &
  \textbf{Success} &
  \textbf{Speedup} \\
SchNet &
  4.82\% &
  1520.54 &
  6.75\% &
  768.60 &
  7.60\% &
  519.72 &
  8.14\% &
  393.38 &
  8.78\% &
  317.71 \\
DimeNet++ &
  10.39\% &
  1518.59 &
  14.03\% &
  766.37 &
  15.85\% &
  511.41 &
  17.34\% &
  385.57 &
  18.20\% &
  308.20 \\
PaiNN &
  39.08\% &
  1464.56 &
  45.40\% &
  742.08 &
  48.82\% &
  497.86 &
  50.64\% &
  373.62 &
  52.03\% &
  299.01 \\
GemNet-OC &
  75.70\% &
  1502.03 &
  84.15\% &
  746.69 &
  87.37\% &
  497.62 &
  88.01\% &
  373.47 &
  89.08\% &
  299.44 \\
GemNet-OC-MD &
  76.77\% &
  1494.44 &
  85.44\% &
  747.21 &
  88.22\% &
  501.47 &
  89.83\% &
  375.80 &
  90.79\% &
  299.95 \\
GemNet-OC-MD-Large &
  80.73\% &
  1455.42 &
  85.97\% &
  736.25 &
  89.29\% &
  492.34 &
  91.11\% &
  370.49 &
  91.86\% &
  296.54 \\
SCN-MD-Large &
  85.12\% &
  1430.68 &
  91.01\% &
  714.02 &
  92.29\% &
  477.07 &
  92.93\% &
  359.51 &
  93.36\% &
  288.46 \\ \midrule
\multicolumn{11}{c}{\textbf{ML+DFT Relaxations (ML+RX)}} \\ \midrule
\textbf{Model} &
  \multicolumn{2}{c}{\textit{\textbf{k=1}}} &
  \multicolumn{2}{c}{\textit{\textbf{k=2}}} &
  \multicolumn{2}{c}{\textit{\textbf{k=3}}} &
  \multicolumn{2}{c}{\textit{\textbf{k=4}}} &
  \multicolumn{2}{c}{\textit{\textbf{k=5}}} \\
\textbf{} &
  \textbf{Success} &
  \textbf{Speedup} &
  \textbf{Success} &
  \textbf{Speedup} &
  \textbf{Success} &
  \textbf{Speedup} &
  \textbf{Success} &
  \textbf{Speedup} &
  \textbf{Success} &
  \textbf{Speedup} \\
SchNet &
  44.54\% &
  47.68 &
  55.46\% &
  24.75 &
  59.74\% &
  17.07 &
  64.45\% &
  13.06 &
  67.45\% &
  10.68 \\
DimeNet++ &
  52.03\% &
  63.92 &
  64.56\% &
  32.90 &
  70.13\% &
  22.22 &
  73.98\% &
  16.74 &
  77.52\% &
  13.48 \\
PaiNN &
  70.24\% &
  97.51 &
  79.44\% &
  48.50 &
  84.05\% &
  32.01 &
  86.83\% &
  23.69 &
  89.08\% &
  19.04 \\
GemNet-OC &
  79.55\% &
  166.33 &
  88.44\% &
  86.65 &
  90.90\% &
  56.63 &
  92.29\% &
  43.62 &
  93.04\% &
  35.30 \\
GemNet-OC-MD &
  78.80\% &
  185.49 &
  87.58\% &
  93.00 &
  91.11\% &
  62.08 &
  92.18\% &
  45.84 &
  93.58\% &
  36.93 \\
GemNet-OC-MD-Large &
  81.80\% &
  201.28 &
  88.44\% &
  100.31 &
  91.54\% &
  67.66 &
  93.15\% &
  51.11 &
  93.68\% &
  41.21 \\
SCN-MD-Large &
  86.72\% &
  185.47 &
  92.83\% &
  96.46 &
  94.43\% &
  64.01 &
  95.07\% &
  47.43 &
  95.50\% &
  37.72 \\ \bottomrule
\end{tabular}%
}
\caption{Model success and speedup results as evaluated against \gls{DFT}-Heuristic across varying $k$ for the OC20-Dense validation set. This evaluation corresponds to a more common community approach.\label{tab:val-dft-heur}}
\end{table*}

%% file: tables/test_subsplits.tex
\begin{table*}[h]
\resizebox{\textwidth}{!}{%
\begin{tabular}{@{}clcccccccccc@{}}
\multicolumn{12}{c}{\gls{OC20D} Test} \\
\toprule
\multirow{2}{*}{\textbf{Split}} &
  \multirow{2}{*}{\textbf{Model}} &
  \multicolumn{5}{c}{\textbf{ML+DFT Singlepoints (ML+SP)}} &
  \multicolumn{5}{c}{\textbf{ML+DFT Relaxations (ML+RX)}} \\ \cmidrule(l){3-12} 
 &
   &
  \textit{\textbf{k=1}} &
  \textit{\textbf{k=2}} &
  \textit{\textbf{k=3}} &
  \textit{\textbf{k=4}} &
  \multicolumn{1}{c|}{\textit{\textbf{k=5}}} &
  \textit{\textbf{k=1}} &
  \textit{\textbf{k=2}} &
  \textit{\textbf{k=3}} &
  \textit{\textbf{k=4}} &
  \textit{\textbf{k=5}} \\ \midrule
\multirow{8}{*}{\textbf{ID}} &
  SchNet &
  2.40\% &
  4.00\% &
  4.00\% &
  4.40\% &
  \multicolumn{1}{c|}{4.80\%} &
  45.20\% &
  55.20\% &
  60.80\% &
  62.40\% &
  65.60\% \\
 &
  DimeNet++ &
  5.60\% &
  6.80\% &
  8.00\% &
  9.20\% &
  \multicolumn{1}{c|}{9.60\%} &
  49.60\% &
  60.00\% &
  64.00\% &
  68.00\% &
  71.20\% \\
 &
  PaiNN &
  36.00\% &
  40.40\% &
  43.60\% &
  44.40\% &
  \multicolumn{1}{c|}{44.80\%} &
  72.80\% &
  78.00\% &
  78.80\% &
  80.80\% &
  81.20\% \\
 &
  GemNet-OC &
  74.40\% &
  79.20\% &
  81.60\% &
  81.60\% &
  \multicolumn{1}{c|}{81.60\%} &
  82.40\% &
  86.40\% &
  87.60\% &
  87.60\% &
  88.00\% \\
 &
  GemNet-OC-MD &
  78.00\% &
  80.00\% &
  80.80\% &
  81.60\% &
  \multicolumn{1}{c|}{82.40\%} &
  81.60\% &
  84.40\% &
  85.60\% &
  86.80\% &
  87.60\% \\
 &
  GemNet-OC-MD-Large &
  78.80\% &
  83.20\% &
  84.80\% &
  85.20\% &
  \multicolumn{1}{c|}{85.60\%} &
  83.60\% &
  86.80\% &
  87.20\% &
  88.40\% &
  88.80\% \\
 &
  SCN-MD-Large &
  77.20\% &
  80.40\% &
  82.40\% &
  84.00\% &
  \multicolumn{1}{c|}{84.80\%} &
  82.40\% &
  86.40\% &
  88.40\% &
  88.40\% &
  89.20\% \\
 &
  eSCN-MD-Large &
  80.40\% &
  83.20\% &
  84.40\% &
  84.80\% &
  \multicolumn{1}{c|}{86.00\%} &
  82.80\% &
  84.80\% &
  86.40\% &
  86.80\% &
  88.40\% \\ \midrule
\multirow{8}{*}{\textbf{OOD-Ads}} &
  SchNet &
  1.63\% &
  1.63\% &
  1.63\% &
  2.04\% &
  \multicolumn{1}{c|}{2.04\%} &
  46.53\% &
  53.88\% &
  58.78\% &
  62.45\% &
  64.08\% \\
 &
  DimeNet++ &
  3.27\% &
  4.08\% &
  4.08\% &
  4.49\% &
  \multicolumn{1}{c|}{5.71\%} &
  46.94\% &
  57.55\% &
  60.82\% &
  63.27\% &
  65.71\% \\
 &
  PaiNN &
  28.16\% &
  33.47\% &
  34.69\% &
  36.33\% &
  \multicolumn{1}{c|}{36.73\%} &
  71.84\% &
  80.00\% &
  84.08\% &
  85.71\% &
  86.53\% \\
 &
  GemNet-OC &
  77.14\% &
  79.59\% &
  81.22\% &
  82.04\% &
  \multicolumn{1}{c|}{83.67\%} &
  84.49\% &
  86.53\% &
  87.76\% &
  87.76\% &
  88.57\% \\
 &
  GemNet-OC-MD &
  82.04\% &
  86.53\% &
  88.57\% &
  88.57\% &
  \multicolumn{1}{c|}{88.57\%} &
  84.08\% &
  89.39\% &
  91.43\% &
  92.24\% &
  93.06\% \\
 &
  GemNet-OC-MD-Large &
  83.27\% &
  86.94\% &
  88.16\% &
  89.39\% &
  \multicolumn{1}{c|}{89.80\%} &
  86.53\% &
  90.20\% &
  92.65\% &
  93.47\% &
  93.88\% \\
 &
  SCN-MD-Large &
  83.67\% &
  88.57\% &
  89.80\% &
  90.20\% &
  \multicolumn{1}{c|}{91.02\%} &
  88.16\% &
  92.65\% &
  93.06\% &
  94.69\% &
  94.69\% \\
 &
  eSCN-MD-Large &
  86.12\% &
  89.80\% &
  91.43\% &
  91.84\% &
  \multicolumn{1}{c|}{91.84\%} &
  87.76\% &
  91.84\% &
  93.06\% &
  93.47\% &
  93.47\% \\ \midrule
\multirow{8}{*}{\textbf{OOD-Cat}} &
  SchNet &
  2.86\% &
  2.86\% &
  2.86\% &
  2.86\% &
  \multicolumn{1}{c|}{2.86\%} &
  45.71\% &
  55.92\% &
  61.63\% &
  63.67\% &
  66.53\% \\
 &
  DimeNet++ &
  4.49\% &
  5.71\% &
  6.53\% &
  6.94\% &
  \multicolumn{1}{c|}{7.35\%} &
  53.06\% &
  63.27\% &
  65.71\% &
  68.98\% &
  73.06\% \\
 &
  PaiNN &
  24.90\% &
  28.98\% &
  31.84\% &
  34.29\% &
  \multicolumn{1}{c|}{35.51\%} &
  75.10\% &
  77.55\% &
  81.22\% &
  84.90\% &
  86.12\% \\
 &
  GemNet-OC &
  64.90\% &
  71.02\% &
  73.06\% &
  75.10\% &
  \multicolumn{1}{c|}{76.73\%} &
  78.78\% &
  82.86\% &
  84.49\% &
  84.49\% &
  86.53\% \\
 &
  GemNet-OC-MD &
  70.61\% &
  76.73\% &
  79.59\% &
  80.00\% &
  \multicolumn{1}{c|}{80.41\%} &
  81.63\% &
  86.53\% &
  88.16\% &
  88.57\% &
  88.98\% \\
 &
  GemNet-OC-MD-Large &
  75.51\% &
  78.78\% &
  81.22\% &
  82.86\% &
  \multicolumn{1}{c|}{84.08\%} &
  82.04\% &
  87.35\% &
  88.98\% &
  90.20\% &
  90.61\% \\
 &
  SCN-MD-Large &
  75.92\% &
  82.04\% &
  83.67\% &
  85.31\% &
  \multicolumn{1}{c|}{86.12\%} &
  82.86\% &
  87.35\% &
  88.57\% &
  90.20\% &
  91.02\% \\
 &
  eSCN-MD-Large &
  79.18\% &
  84.49\% &
  86.53\% &
  86.53\% &
  \multicolumn{1}{c|}{86.94\%} &
  82.86\% &
  87.76\% &
  88.57\% &
  88.57\% &
  90.20\% \\ \midrule
\multirow{8}{*}{\textbf{OOD-Both}} &
  SchNet &
  2.41\% &
  2.41\% &
  2.41\% &
  2.81\% &
  \multicolumn{1}{c|}{2.81\%} &
  41.37\% &
  53.82\% &
  59.44\% &
  63.45\% &
  65.86\% \\
 &
  DimeNet++ &
  7.23\% &
  7.63\% &
  8.84\% &
  8.84\% &
  \multicolumn{1}{c|}{9.24\%} &
  51.41\% &
  59.04\% &
  63.45\% &
  65.86\% &
  69.48\% \\
 &
  PaiNN &
  26.51\% &
  30.92\% &
  32.93\% &
  34.14\% &
  \multicolumn{1}{c|}{36.14\%} &
  69.48\% &
  75.10\% &
  79.52\% &
  83.13\% &
  85.14\% \\
 &
  GemNet-OC &
  73.90\% &
  77.51\% &
  81.12\% &
  81.93\% &
  \multicolumn{1}{c|}{81.93\%} &
  82.73\% &
  86.75\% &
  88.35\% &
  88.35\% &
  88.76\% \\
 &
  GemNet-OC-MD &
  75.10\% &
  79.92\% &
  81.53\% &
  83.13\% &
  \multicolumn{1}{c|}{83.94\%} &
  82.33\% &
  85.94\% &
  87.95\% &
  89.16\% &
  89.96\% \\
 &
  GemNet-OC-MD-Large &
  79.12\% &
  82.73\% &
  83.53\% &
  85.14\% &
  \multicolumn{1}{c|}{86.35\%} &
  80.72\% &
  87.15\% &
  89.56\% &
  91.57\% &
  93.17\% \\
 &
  SCN-MD-Large &
  76.71\% &
  82.73\% &
  84.74\% &
  86.35\% &
  \multicolumn{1}{c|}{86.35\%} &
  83.13\% &
  88.76\% &
  89.16\% &
  90.36\% &
  90.76\% \\
 &
  eSCN-MD-Large &
  82.33\% &
  84.74\% &
  87.15\% &
  88.35\% &
  88.35\% &
  83.13\% &
  87.95\% &
  89.16\% &
  90.36\% &
  90.36\% \\ \bottomrule
\end{tabular}%
}
\label{tab:w2b-splits}
\caption{Success rates evaluated on \gls{DFT}-Heur+Rand across the different in-domain and out-of-domain subsplits for the \gls{OC20D} test set. Results reported for both ML+SP and ML+RX strategies across different $k$ values.\label{tab:test-splits}}
\end{table*}

%% file: tables/val_subsplits.tex
\begin{table*}[h]
\resizebox{\textwidth}{!}{%
\begin{tabular}{@{}clccccc|ccccc@{}}
\multicolumn{12}{c}{\gls{OC20D} Validation} \\
\toprule
\multirow{2}{*}{\textbf{Split}} &
\multirow{2}{*}{\textbf{Model}} &
\multicolumn{5}{c|}{\textbf{ML+DFT Single-points (ML+SP)}} &
\multicolumn{5}{c}{\textbf{ML+DFT Relaxations (ML+RX)}} \\ \cmidrule(l){3-12} 
&
&
\textit{\textbf{k=1}} &
\textit{\textbf{k=2}} &
\textit{\textbf{k=3}} &
\textit{\textbf{k=4}} &
\textit{\textbf{k=5}} &
\textit{\textbf{k=1}} &
\textit{\textbf{k=2}} &
\textit{\textbf{k=3}} &
\textit{\textbf{k=4}} &
\textit{\textbf{k=5}} \\ \midrule
\multirow{7}{*}{\textbf{ID}}       & SchNet                      & 1.64\%  & 2.87\%  & 3.69\%  & 3.69\%  & 3.69\%  & 37.70\% & 50.41\% & 54.92\% & 60.25\% & 62.30\% \\
& DimeNet++                   & 3.69\%  & 4.51\%  & 4.92\%  & 6.56\%  & 7.38\%  & 46.31\% & 55.33\% & 60.66\% & 65.57\% & 67.62\% \\
& PaiNN                       & 31.97\% & 39.34\% & 43.03\% & 45.08\% & 46.31\% & 60.66\% & 72.54\% & 75.82\% & 78.28\% & 79.92\% \\
& GemNet-OC                   & 76.23\% & 81.15\% & 84.84\% & 85.66\% & 86.48\% & 74.18\% & 84.84\% & 90.57\% & 92.62\% & 92.62\% \\
& GemNet-OC-MD                & 71.31\% & 81.15\% & 84.02\% & 86.48\% & 87.30\% & 73.77\% & 84.84\% & 86.89\% & 88.11\% & 88.52\% \\
& GemNet-OC-MD-Large          & 77.87\% & 82.79\% & 85.25\% & 87.30\% & 87.30\% & 77.05\% & 83.61\% & 86.07\% & 88.52\% & 88.52\% \\
& SCN-MD-Large                & 81.15\% & 85.66\% & 87.30\% & 88.52\% & 88.52\% & 82.38\% & 86.48\% & 87.70\% & 88.93\% & 89.34\% \\ \midrule
\multirow{7}{*}{\textbf{OOD-Ads}}  & SchNet                      & 4.51\%  & 6.56\%  & 6.56\%  & 6.97\%  & 7.38\%  & 37.70\% & 48.77\% & 54.10\% & 59.43\% & 60.66\% \\
& DimeNet++                   & 5.74\%  & 7.79\%  & 9.84\%  & 11.48\% & 12.70\% & 45.49\% & 56.15\% & 63.93\% & 67.62\% & 71.31\% \\
& PaiNN                       & 29.51\% & 36.07\% & 38.93\% & 40.57\% & 41.80\% & 63.52\% & 72.95\% & 77.05\% & 79.51\% & 81.56\% \\
& GemNet-OC                   & 68.44\% & 79.51\% & 84.02\% & 84.43\% & 86.07\% & 74.18\% & 81.97\% & 85.66\% & 86.48\% & 88.93\% \\
& GemNet-OC-MD                & 70.90\% & 81.15\% & 84.43\% & 86.07\% & 86.89\% & 71.72\% & 81.15\% & 86.07\% & 88.11\% & 88.93\% \\
& GemNet-OC-MD-Large          & 74.18\% & 83.20\% & 85.66\% & 88.52\% & 89.75\% & 78.28\% & 84.02\% & 88.11\% & 90.98\% & 91.39\% \\
& SCN-MD-Large                & 77.87\% & 84.02\% & 86.89\% & 88.11\% & 88.93\% & 79.10\% & 86.89\% & 88.93\% & 90.16\% & 90.98\% \\ \midrule
\multirow{7}{*}{\textbf{OOD-Cat}}  & SchNet                      & 1.68\%  & 2.10\%  & 2.94\%  & 4.20\%  & 4.62\%  & 34.87\% & 44.12\% & 50.00\% & 54.62\% & 57.56\% \\
& DimeNet++          & 7.14\%  & 10.08\% & 10.50\% & 11.34\% & 11.76\% & 40.34\% & 54.20\% & 57.56\% & 61.76\% & 65.97\% \\
& PaiNN              & 25.63\% & 31.93\% & 35.29\% & 37.39\% & 38.24\% & 67.23\% & 76.05\% & 79.83\% & 84.03\% & 86.13\% \\
& GemNet-OC          & 70.59\% & 78.99\% & 82.35\% & 83.19\% & 85.29\% & 77.73\% & 86.97\% & 89.08\% & 89.50\% & 90.76\% \\
& GemNet-OC-MD       & 71.43\% & 80.67\% & 84.03\% & 84.45\% & 85.71\% & 74.79\% & 82.35\% & 86.97\% & 89.50\% & 92.02\% \\
& GemNet-OC-MD-Large & 70.17\% & 76.05\% & 82.35\% & 84.45\% & 84.87\% & 76.47\% & 82.77\% & 86.55\% & 88.66\% & 89.50\% \\
& SCN-MD-Large                & 80.67\% & 89.08\% & 89.92\% & 90.76\% & 91.18\% & 83.61\% & 92.02\% & 94.54\% & 95.38\% & 95.38\% \\ \midrule
\multirow{7}{*}{\textbf{OOD-Both}} & SchNet                      & 3.24\%  & 4.05\%  & 4.05\%  & 4.05\%  & 4.45\%  & 30.77\% & 40.89\% & 45.34\% & 47.77\% & 51.82\% \\
& DimeNet++          & 4.86\%  & 8.10\%  & 10.12\% & 10.93\% & 11.34\% & 41.70\% & 52.63\% & 61.13\% & 63.56\% & 66.80\% \\
& PaiNN              & 22.67\% & 27.13\% & 29.55\% & 31.58\% & 31.98\% & 55.47\% & 63.16\% & 70.45\% & 74.49\% & 77.33\% \\
& GemNet-OC          & 59.92\% & 69.64\% & 72.06\% & 72.87\% & 74.09\% & 68.42\% & 78.95\% & 82.19\% & 84.21\% & 85.43\% \\
& GemNet-OC-MD       & 61.54\% & 70.04\% & 74.90\% & 76.11\% & 77.73\% & 68.83\% & 77.33\% & 80.57\% & 82.59\% & 84.62\% \\
& GemNet-OC-MD-Large & 70.45\% & 76.52\% & 79.76\% & 81.38\% & 82.19\% & 72.87\% & 81.78\% & 83.81\% & 84.62\% & 86.23\% \\
& SCN-MD-Large                & 71.66\% & 78.54\% & 81.38\% & 82.19\% & 82.59\% & 77.33\% & 82.59\% & 86.23\% & 86.23\% & 87.04\% \\ \bottomrule
\end{tabular}%
}
\caption{Success rates evaluated on \gls{DFT}-Heur+Rand across the different in-domain and out-of-domain subsplits for the \gls{OC20D} validation set. Results reported for both ML+SP and ML+RX strategies across different $k$ values.\label{tab:val-splits}}
\end{table*}

%% file: tables/compute.tex
\begin{table*}[h]
\resizebox{0.5\textwidth}{!}{%
\begin{tabular}{@{}lccc@{}}
\toprule
& \textbf{ML RX}  & \textbf{DFT SP} & \textbf{DFT RX} \\
\textbf{Model}     & \textbf{(GPU-hrs)} & \textbf{(CPU-hrs)} & \textbf{(CPU-hrs)} \\ \midrule
SchNet             & 24.2            & 2,199.16              & 51,989.30             \\
DimeNet++          & 249.2           & 2,576.00              & 54,785.46             \\
PaiNN              & 60.4            & 2,225.63              & 38,409.31             \\
GemNet-OC          & 133.0           & 2,824.25              & 25,073.62             \\
GemNet-OC-MD       & 133.0           & 2,441.12              & 26,411.08             \\
GemNet-OC-MD-Large & 638.3          & 2,448.27              & 19,265.97             \\
SCN-MD-Large       & 1129.2          & 2,645.39              & 17,313.90             \\ \midrule
DFT Heuristic      & -               & -                     & 806,351.19            \\
DFT Random         & -               & -                     & 1,096,396.77          \\ \bottomrule
\end{tabular}%
}
\caption{Total compute time associated with \gls{ML} relaxations, \gls{DFT} singlepoints (SP) and relaxations (RX) on best $k=5$ \gls{ML} predictions for the \gls{OC20D} validation set. Baseline \gls{DFT} Heuristic and Random ground truths are also shown for reference.\label{tab:compute}}
\end{table*}

%% file: tables/alt_speedup.tex
\begin{table*}[]
\resizebox{0.9\textwidth}{!}{%
\begin{tabular}{@{}lcccc|cccc@{}}
\multicolumn{9}{c}{Alternative DFT Speedup} \\ \midrule
\multirow{3}{*}{\textbf{Model}} & \multicolumn{4}{c|}{\textit{k=1}} & \multicolumn{4}{c}{\textit{k=5}} \\ \cmidrule(l){2-9} 
 & \multicolumn{2}{c}{\textbf{ML+SP}} & \multicolumn{2}{c|}{\textbf{ML+RX}} & \multicolumn{2}{c}{\textbf{ML+SP}} & \multicolumn{2}{c}{\textbf{ML+RX}} \\
 & without ML & with ML & without ML & with ML & without ML & with ML & without ML & with ML \\ \midrule
SchNet & 4326.08 & 4100.65 & 182.99 & 182.57 & 865.22 & 855.81 & 36.60 & 36.58 \\
DimeNet++ & 3693.23 & 2489.08 & 173.65 & 169.79 & 738.65 & 673.48 & 34.73 & 34.57 \\
PaiNN & 4274.63 & 3763.54 & 247.69 & 245.76 & 854.93 & 832.32 & 49.54 & 49.46 \\
GemNet-OC & 3368.59 & 2726.64 & 379.43 & 369.63 & 673.72 & 643.42 & 75.89 & 75.49 \\
GemNet-OC-MD & 3897.29 & 3062.98 & 360.22 & 351.37 & 779.46 & 739.19 & 72.04 & 71.68 \\
GemNet-OC-MD-Large & 3885.90 & 1686.86 & 493.81 & 423.63 & 777.18 & 616.45 & 98.76 & 95.59 \\
SCN-MD-Large & 3596.35 & 1147.45 & 549.49 & 414.37 & 719.27 & 504.10 & 109.90 & 103.17 \\ \bottomrule
\end{tabular}%
}
\caption{Alternative speedup metric as computed by total runtime across all models on the \gls{OC20D} validation set. Speedup is computed with and without factoring in \gls{ML} runtime to compare results. Results are evaluated for both ML+SP and ML+RX strategies at $k=1$ and $k=5$.\label{tab:alt-speedup}}
\end{table*}

%% file: tables/dedup.tex
\begin{table*}[h]
\resizebox{0.4\textwidth}{!}{%
\begin{tabular}{@{}cccccc@{}}
\toprule
\multirow{2}{*}{$\Delta E$} & \multicolumn{5}{c}{\textbf{Success}} \\
 & \textit{k=1} & \textit{k=2} & \textit{k=3} & \textit{k=4} & \textit{k=5} \\ \midrule
0 & 77.80\% & 84.28\% & 86.33\% & 87.36\% & 87.77\% \\
1.00E-09 & 77.80\% & 84.28\% & 86.33\% & 87.36\% & 87.77\% \\
0.005 & 77.80\% & 84.28\% & 86.43\% & 87.36\% & 87.98\% \\
0.01 & 77.80\% & 84.69\% & 86.43\% & 87.46\% & 87.98\% \\
0.02 & 77.80\% & 83.25\% & 84.79\% & 85.61\% & 86.13\% \\ \bottomrule
\end{tabular}
}
\caption{Deduplication results with SCN-MD-Large ML+SP under different $\Delta E$ cluster thresholds. Success rates computed against the \gls{DFTHR} ground truth on the \gls{OC20D} validation set.\label{tab:dedup}}
\end{table*}

%% file: tables/heur_xrand.tex
\begin{table}[h]
\begin{tabular}{@{}ccc@{}}
\toprule
\textbf{+\% Random} & \textbf{Success Rate} & \textbf{Speedup} \\ \midrule
0\% & 71.12\% & 2.87 \\
10\% & 79.45\% & 2.45 \\
20\% & 86.33\% & 2.11 \\
30\% & 88.90\% & 1.86 \\
40\% & 91.06\% & 1.66 \\
50\% & 93.42\% & 1.49 \\
60\% & 94.86\% & 1.36 \\
70\% & 96.20\% & 1.25 \\
80\% & 98.36\% & 1.16 \\
90\% & 99.08\% & 1.08 \\
100\% & 100.00\% & 1.00 \\ \bottomrule
\end{tabular}
\caption{Success rate and speedup of varying proportions of random configurations added to the heuristics. \gls{DFTH} corresponds to the 0.0\% data point, and \gls{DFTHR} corresponds to the ground truth used throughout the paper. Results reported for the \gls{OC20D} validation set.\label{tab:heur+xrand}}
\end{table}

%% file: tables/compare-rand-heur.tex
\begin{table*}[h]
\resizebox{0.5\textwidth}{!}{%
\begin{tabular}{@{}cccccc@{}}
\toprule
\multirow{2}{*}{Configuration type} & \multicolumn{5}{c}{\textbf{Success}} \\
 & \textit{k=1} & \textit{k=2} & \textit{k=3} & \textit{k=4} & \textit{k=5} \\ \midrule
Heuristic ML, GT-both  & 57.04\% & 60.64\% & 61.36\% & 61.87\% & 62.18\% \\
Random ML, GT-both  & 73.48\% & 79.03\% & 81.91\% & 82.73\% & 82.94\% \\
Heuristic ML, GT-heuristic  & 77.94\% & 83.30\% & 84.05\% & 84.69\% & 85.12\% \\
Random ML, GT-random  & 78.11\% & 83.61\% & 86.41\% & 87.24\% & 87.45\% \\
\bottomrule
\end{tabular}
}
\caption{Comparing random and heuristic configurations on the \gls{OC20D} validation set.. Heuristic ML represents using the \algo{} algorithm but only on heuristic initial configurations, and Random ML uses only random configurations. GT-both considers both heuristic and random for ground truth (which was done for the main results), and GT-heuristic and GT-random mean that ground truth only uses heuristic or random configurations, respectively. Results show that removing random configurations decreases the success rates more. When switching to GT-heuristic, Heuristic ML becomes competitive again, indicating that random configurations helps both AdsorbML and ground truth.\label{tab:comprandheur}
}
\end{table*}

%% file: tables/random-and-worst.tex
\begin{table*}[h]
\resizebox{0.5\textwidth}{!}{%
\begin{tabular}{@{}cccccc@{}}
\toprule
\multirow{2}{*}{Binding site selection} & \multicolumn{5}{c}{\textbf{Success}} \\
 & \textit{k=1} & \textit{k=2} & \textit{k=3} & \textit{k=4} & \textit{k=5} \\ \midrule
Best k   & 77.80\% & 84.28\% & 86.33\% & 87.36\% & 87.77\% \\
Random k & 20.90\% & 31.86\% & 40.08\% & 45.53\% & 50.15\% \\
Worst k  & 1.85\%  & 3.39\%  & 4.32\%  & 5.14\%  & 6.27\% \\

\bottomrule
\end{tabular}
}
\caption{Comparison against random and worst ranking baselines for SCN-MD-Large single-points on the \gls{OC20D} validation set. ``Best k'' refers to the regular algorithm, with the same results as in Table \ref{tab:main-dft-heur_rand}. For ``Random k'', we choose a random set of k placements and averaged success rates across three seeds. For ``Worst k'', we choose the placements with highest ML predicted energies rather than lowest. As expected, random performs badly and choosing high energy placements performs the worst.\label{tab:rand-and-worst-k}}
\end{table*}

%% file: tables/anomalies.tex
\begin{table*}[h]
\begin{tabular}{@{}lcccc@{}}
\toprule
\textbf{Model} & \textbf{Dissociation} & \textbf{Desorption} & \textbf{Surface mismatch} & \textbf{Total} \\ \midrule
SchNet & 10,287 & 4,183 & 40,639 & \textbf{48,546} \\
DimeNet++ & 9,706 & 6,389 & 17,491 & \textbf{29,913} \\
PaiNN & 8,273 & 5,211 & 8,656 & \textbf{20,539} \\
GemNet-OC & 8,944 & 4,815 & 10,105 & \textbf{22,143} \\
GemNet-OC-MD & 8,781 & 5,019 & 9,526 & \textbf{21,676} \\
GemNet-OC-MD-Large & 8,871 & 4,693 & 10,000 & \textbf{21,829} \\
SCN-MD-Large & 8,524 & 4,972 & 9,048 & \textbf{20,860} \\ \midrule
\gls{DFTHR} & 9,075 & 3,491 & 8,407 & \textbf{19,432} \\ \bottomrule
\end{tabular}
\caption{Breakdown of relaxations removed for violating proposed constraints for \gls{ML} and \gls{DFT} ground truth relaxations. Note, a system may have more than one violation type, hence the total may not correspond to the sum across all types. Counts reported on the validation set. \label{tab:anomalies}}
\end{table*}

%% file: sections/changelog.tex
\section{Changelog}

This section tracks the changes to this document since the original release.

\noindent \textbf{v1}. Initial version.

\noindent \textbf{v2}.
\begin{itemize}[noitemsep,topsep=0pt]
    \item Updated DFT-Heuristic and DFT-Random total compute times, ignoring systems in which were run but ignored from evaluation.
    \item Updated all speedup numbers as a result of the updated DFT compute times.
    \item Updated the OC20-Dense dataset statistics, ignoring systems that were removed from evaluation due to problematic inputs.
\end{itemize}

\noindent v3. Published in \textit{npj Comput. Mater.}
\begin{itemize}[noitemsep,topsep=0pt]
    \item Introduced the OC20-Dense Test set, curated in a similar manner to the previous validation set.
    \item Evaluated all models on the new OC20-Dense Test set, updating all evaluation metrics across the manuscript's tables and figures.
    \item Included eSCN-MD-Large, a more recent state-of-the-art OC20 model, to the set of models for OC20-Dense evaluation.
\end{itemize}